\documentclass[12pt]{iopart}

\usepackage{iopams,subeqn}
\usepackage{color}

\usepackage{graphicx}

\newcommand{\oneminus}[1]{{#1}^{\prime}}
\newcommand{\dfull}{{p}}
\newcommand{\dbulk}{{q}}
\newcommand{\ddetr}{{r}}
\newcommand{\ddlim}{{s}}
\newcommand{\eqref}{\eref}
\newcommand{\z}{{z}}

\newcommand{\Omegamedio}{\frac{\Omega_\mathrm{A} + \Omega_\mathrm{D}}{2}}

\begin{document}
\title[TASEP with unbalanced Langmuir kinetics]{Unbalanced Langmuir kinetics affects TASEP dynamical transitions: mean-field theory}
\author{D Botto$^{1,2}$, A Pelizzola$^{1,3}$,
M Pretti$^{1,4}$ and M Zamparo$^1$}
\address{$^1$ Dipartimento Scienza Applicata e Tecnologia, Politecnico di Torino, Corso Duca degli Abruzzi 24, 10129 Torino, Italy}
\address{$^2$ Informatica System s.r.l., Piazzetta del Borgo 1, 12080 Vicoforte (CN), Italy}
\address{$^3$ INFN, Sezione di Torino, via Pietro Giuria 1, 10125 Torino, Italy}
\address{$^4$ Consiglio Nazionale delle Ricerche - Istituto dei Sistemi Complessi (CNR-ISC), Via dei Taurini 19, 00185 Roma, Italy}
\eads{\mailto{davide.botto@polito.it}, \mailto{alessandro.pelizzola@polito.it}, \mailto{marco.pretti@polito.it}, \mailto{marco.zamparo@polito.it}}
\begin{abstract}
In a previous study we developed a mean-field theory of dynamical transitions for the \emph{totally-asymmetric simple-exclusion process} (TASEP) with open boundaries and Langmuir kinetics, in the so-called \emph{balanced} regime, characterized by equal binding and unbinding rates. 
Here we show that simply including the possibility of unbalanced rates gives rise to  an unexpectedly richer dynamical phase diagram. 
In particular, the current work predicts an unusual type of dynamical transition, which exhibits certain similarities with first-order phase transitions of equilibrium systems. 
We also point out that different types of dynamical transition are accompanied by different structural changes in the (mean-field) relaxation spectrum. 
\end{abstract}
\maketitle

\section{Introduction}
\label{sec:intro}

The \textit{totally-asymmetric simple-exclusion process} (TASEP) is a paradigmatic model of stochastic transport, with a variety of physical applications and mathematical connections, ranging from molecular biophysics to vehicular traffic~\cite{ChouMallickZia11,SchadschneiderChowdhuryNishinari11}, up to the Kardar-Parisi-Zhang universality and random-matrix theory~\cite{Lazarescu15,KriecherbauerKrug10}. 
In its simplest form, the model is defined on a one-dimensional lattice, whose nodes can be occupied by at most one particle, and where each particle can hop to the adjacent node, provided the latter is empty. 
The model is called \emph{totally asymmetric} if hopping occurs exclusively in one direction. 
In the \emph{open} TASEP, which we consider here, particles are injected at one end of the lattice and extracted at the opposite end, at given rates. 
This model is at the core of a lot of fundamental studies in non-equilibrium statistical mechanics~\cite{ChouMallickZia11,SchadschneiderChowdhuryNishinari11,Lazarescu15}, because on the one hand it is simple enough that several properties can be worked out exactly (the steady-state solution dates back to the 1990s~\cite{DerridaDomanyMukamel92,SchutzDomany93,Derrida-etal93,Derrida98}) and on the other hand it exhibits non-trivial (non-equilibrium) steady states and a number of phase transitions among them, controlled by the injection and extraction rates (boundary-induced phase transitions). 

One possible generalization of the TASEP is the so-called TASEP with Langmuir kinetics (TASEP-LK), in which particles can also bind to an empty node or unbind from an occupied one. 
The latter model was first introduced in \cite{WillmannSchutzChallet02}, with a specific application in econophysics, but it is thought to be relevant even in very different contexts such as, for instance, that of intracellular transport~\cite{NishinariOkadaSchadschneiderChowdhury05,GreulichSchadschneider09}. 
The Langmuir kinetics does not allow for an exact solution of the model, which has thus been studied by mean-field approximations, hydrodynamic equations and numerical simulations~\cite{ParmeggianiFranoschFrey03,Popkov03,Evans03,ParmeggianiFranoschFrey04}, reaching a rather complete description of the physics underlying the steady-state behaviour. 

As far as the ordinary TASEP is concerned (without Langmuir kinetics), a purely dynamical phase transition has also been discovered by de Gier and Essler via exact methods~\cite{deGierEssler05,deGierEssler06,deGierEssler08}. 
Such a transition emerges as a singularity in the relaxation rate of the system, without affecting any other steady-state property. 
The transition separates a ``normal'' dynamical regime, where the relaxation rate depends on a given control parameter (the injection/extraction rate, according to the region of the phase diagram) from a ``saturated'' one, in which the relaxation rate reaches a maximum and remains constant thereafter.
Mostly relying on mean-field-like techniques, some recent works~\cite{BottoPelizzolaPretti18,BPPZ18,PelizzolaPrettiPuccioni19} provide evidence that similar dynamical transitions should occur as well in different generalizations of the TASEP, with Langmuir kinetics or with local particle interactions. 
For the pure TASEP, the same kind of approximations~\cite{PelizzolaPretti17} predict a dynamical transition line in good qualitative agreement with the exact one, slightly improving previous approaches based either on the domain-wall theory~\cite{KolomeiskySchutzKolomeiskyStraley98,DudzinskiSchutz00,NagyAppertSanten02} or on the viscous Burgers equation~\cite{ProemeBlytheEvans11}. 
It is also interesting to note that, among the cited papers, \cite{BottoPelizzolaPretti18} and \cite{PelizzolaPretti17} explicitly trace a connection between the onset of dynamical transitions and structural changes in the (mean-field) relaxation spectrum. 

In this paper we extend the mean-field theory for the TASEP-LK presented in~\cite{BPPZ18}, in order to deal with so-called \emph{unbalanced} Langmuir kinetics, that is with unequal binding and unbinding rates. 
Unexpectedly, we find out that so simple a generalization yields a much richer dynamical phase diagram, including a novel type of dynamical transition, that recalls, in different aspects, the usual first-order phase transitions for equilibrium systems. 
We also show that this type of transition is associated to a specific behaviour of the relaxation spectrum, which differs from the one accompanying the ``ordinary'' dynamical transition. 

The paper is organized as follows. 
In section~\ref{sec:model} we define the model and introduce the mean-field theory. 
In section~\ref{sec:phase} we report some known features of the steady-state, namely phase diagram and density profiles, that are of interest for the current work. 
Section~\ref{sec:dyn_trans} is the central one, describing our original contributions. 
We recap our findings and draw some conclusions in section~\ref{sec:conclusion}.
The technical details are reported in two appendices.

\section{The model and the mean-field theory}
\label{sec:model}

In this section we present the model and give a brief account of the mean field theory, which has been previously reported in~\cite{BPPZ18}.

The TASEP dynamics is defined as usual. 
We have a one-dimensional lattice of $N$ nodes, where each node may be either empty or occupied by at most one particle, and each particle hops to the rightward nearest-neighbour node (provided the latter is empty) with unit rate. 
Moreover, particles are injected at the leftmost node (provided it is empty) with rate $\alpha$, and extracted from the rightmost node (provided it is occupied) with rate $\beta$. 
In the TASEP-LK, independent binding and unbinding processes (Langmuir kinetics) also take place at each node. 
In particular a particle can bind to an empty node with rate $\omega_\mathrm{A} \equiv \Omega_\mathrm{A}/(N+1)$ (attachment rate) or unbind from an occupied node with rate $\omega_\mathrm{D} \equiv \Omega_\mathrm{D}/(N+1)$ (detachment rate), where the ``macroscopic'' rates $\Omega_\mathrm{A}$ and $\Omega_\mathrm{D}$ are independent of the system size. 
Let us recall that the inverse-of-$N$ scaling is the only physically interesting case \cite{ParmeggianiFranoschFrey03}, in that a competition can be established between the injection/extraction processes and the Langmuir kinetics, even in the limit of large~$N$. 

The theory can be summarized as follows. 
Let ${n = 1,\dots,N}$ be node labels from left to right and let $\dfull_{n}(t)$ denote the \emph{occupation probability} (which we shall also call \emph{local density}) of node $n$ at time $t$. 
The injection and extraction processes can be represented respectively by two extra nodes ${n=0}$ and ${n=N+1}$ of fixed densities\footnote
{
	Here we avoid the term \emph{probabilities}, because in principle $\dfull_{0}$ may be larger than $1$ (if ${\alpha > 1}$), whereas $\dfull_{N+1}$ may be less than $0$ (if ${\beta > 1}$). 
	In \eqref{eq:p_right} we also introduce the notation ${\oneminus{x} \equiv 1-x}$.
} 
\begin{subequations} 
\label{eq:p_boundaries} 
\begin{eqnarray}
\dfull_{0} & = \alpha 
\, , 
\label{eq:p_left}
\\
\dfull_{N+1} & = 1-\beta \equiv \oneminus{\beta}
\, . 
\label{eq:p_right}
\end{eqnarray}
\end{subequations} 
In general, the time evolution of local densities must obey a set of continuity equations. 
Taking into account both hopping and Langmuir kinetics, such equations read
\begin{equation}
\dot{\dfull}_{n}(t)
= {J}_{n-1}(t) - {J}_{n}(t)
+ \omega_\mathrm{A} \oneminus{\dfull_{n}}(t)
- \omega_\mathrm{D} \dfull_{n}(t)
\qquad n = 1,\dots,N
\, ,
\label{eq:p_time_evolution}
\end{equation}
where ${J}_{n}(t)$ denotes the probability current from node $n$ to ${n+1}$, i.e.~(the hopping rate being unity by construction) the probability that node $n$ is occupied and node ${n+1}$ is empty. 
The mean-field approximation amounts to neglect correlations, that is to impose 
\begin{equation}
{J}_{n}(t) \equiv \dfull_{n}(t) \, \oneminus{\dfull_{n+1}}(t)
\qquad n = 0,\dots,N
\, ,
\label{eq:J_vs_p}
\end{equation}
thus closing the set of equations. 
Steady-state equations can then be obtained imposing that in \eqref{eq:p_time_evolution} the time derivatives $\dot{\dfull}_{n}(t)$ vanish, which yields
\begin{equation}
\dfull_{n} \left( \oneminus{\dfull_{n+1}} + \omega_\mathrm{D} \right)
= \left( \dfull_{n-1} + \omega_\mathrm{A} \right) \oneminus{\dfull_{n}}
\qquad n = 1,\dots,N
\, ,
\label{eq:p_steady_state}
\end{equation}
where (as we shall do from now on) steady-state occupation probabilities are simply denoted by dropping the time variable. 
These last equations can be easily solved numerically by a fixed-point method \cite{BPPZ18}. 
In order to investigate the relaxation process, we linearize \eqref{eq:p_time_evolution}-\eqref{eq:J_vs_p} around the steady state, which naturally leads to a system of (first-order, linear) ordinary differential equations, characterized by a tridiagonal coefficient matrix (relaxation matrix). 
Denoting by $\lambda$ a generic eigenvalue of the relaxation matrix, and by ${{v}_{1},\dots,{v}_{N}}$ the components of the corresponding eigenvector, the eigenvalue problem can be written as \cite{BPPZ18}
\begin{equation}
{a}_{n} {v}_{n} - \dfull_{n} {v}_{n+1} - \oneminus{\dfull_{n}} {v}_{n-1}
= \lambda {v}_{n}
\qquad n = 1,\dots,N
\, 
\end{equation}
with 
\begin{equation}
{a}_{n} \equiv \oneminus{\dfull_{n+1}} + \dfull_{n-1}  
+ \omega_\mathrm{A} + \omega_\mathrm{D}
\label{eq:definizione_a}
\end{equation}
and with boundary conditions 
\begin{equation}
{v}_{0} = {v}_{N+1} = 0
\, .
\end{equation}
Throughout this paper, we shall mainly be interested in the smallest $\lambda$, i.e.~the slowest relaxation rate, being the one relevant at long times. 
Let us remark the fact that, even though the relaxation matrix is non-symmetric, its off-diagonal entries never change sign, which allows one to perform a similarity transformation to a (tridiagonal) symmetric matrix. 
In formulae, defining 
\begin{equation}
{u}_{n} \equiv {v}_{n} 
\prod_{k=0}^{n-1}
\sqrt{\frac{\dfull_{k}}{\oneminus{\dfull_{k+1}}}} 
\qquad n = 0,\dots,N+1 
\label{eq:similarity_transformation}
\end{equation}
(where it is understood that the product is $1$ for ${n = 0}$), we get
\begin{equation}
{a}_{n} {u}_{n} 
- \sqrt{\dfull_{n} \oneminus{\dfull_{n+1}}} \, {u}_{n+1} 
- \sqrt{\dfull_{n-1} \oneminus{\dfull_{n}}} \, {u}_{n-1}
= \lambda {u}_{n}
\qquad n = 1,\dots,N
\label{eq:eigenvalue_equation_symmetrized}
\end{equation}
with the analogous boundary conditions
\begin{equation}
{u}_{0} = {u}_{N+1} = 0
\, .
\end{equation}

\section{Static phase diagram and density profiles}
\label{sec:phase}

The static phase diagram of the TASEP-LK has been thoroughly investigated by Parmeggiani, Franosch and Frey~\cite{ParmeggianiFranoschFrey03,ParmeggianiFranoschFrey04}. 
As for the pure TASEP, it turns out that the mean field approximation correctly reproduces the bulk behavior of the system, and thence the static phase diagram~\cite{ParmeggianiFranoschFrey04}. 
In this section we report the main features of the latter, in order to put our results in the proper framework. 

\begin{figure}
	\centering
	\resizebox{106mm}{!}{\includegraphics*{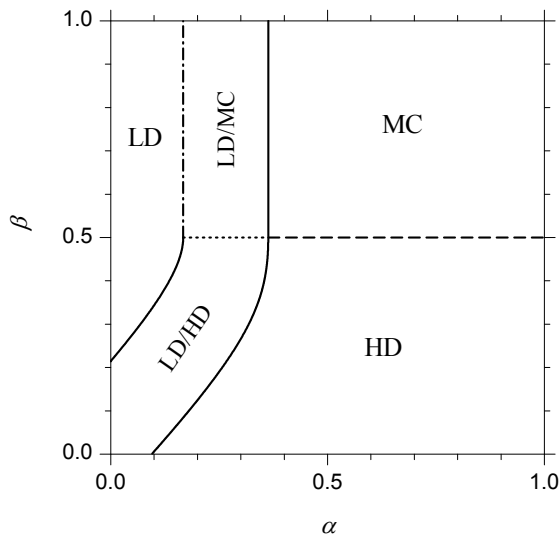}}
	\caption
	{
		Static phase diagram in the $\alpha$-$\beta$ plane for ${\Omega_\mathrm{D} = 0.1}$ and ${\Omega_\mathrm{A} = 0.2}$. 
		Phase labels and different transition line types are explained in the text. 
		Multiple labels denote coexistence regions. 
	}
	\label{fig:phase_diagram}
\end{figure}

First of all, let us recall that, with respect to the pure TASEP (or the balanced TASEP-LK as well), the particle-hole symmetry takes a more general form, namely\footnote{The symmetry involving occupation probabilities applies more generally to occupation numbers.} ${\Omega_\mathrm{A} \leftrightarrow \Omega_\mathrm{D}}$, ${\alpha \leftrightarrow \beta}$, ${\dfull_{n} \leftrightarrow \oneminus{\dfull_{N+1-n}}}$. 
As a consequence, the phase diagram is no longer symmetric under simple exchange of $\alpha$ and $\beta$, but anyway, fixing $\Omega_\mathrm{A}$ and $\Omega_\mathrm{D}$, the analysis can be restricted to ${\Omega_\mathrm{A} > \Omega_\mathrm{D}}$, as we shall assume from now on. 
Figure~\ref{fig:phase_diagram} displays a typical phase diagram for such a case.
One can still observe three phases characterizing the pure TASEP, as usual named high-density (HD), low-density (LD) and maximal-current (MC) phase.  
These phases can be regarded as continuous evolutions of the respective pure-TASEP phases (obviously recovered for ${\Omega_\mathrm{A} = \Omega_\mathrm{D} = 0}$), since they respectively retain their relevant features. 
In particular, in the HD (resp.~LD) phase the bulk behaviour is completely determined by the extraction rate $\beta$ (injection rate $\alpha$) alone, whereas $\alpha$ (resp.~$\beta$) just plays a role in the formation of the boundary layer. 
Conversely, the bulk behaviour of the MC phase is absolutely unaffected by either $\alpha$ and $\beta$ (so that in \cite{ParmeggianiFranoschFrey04} it is denoted as the Meissner phase), whereas both parameters determine only the boundary layers. 
So far, the main difference with respect to the pure TASEP is the lack of uniformity in the bulk density and current profiles, which is naturally induced by Langmuir kinetics. 
In particular, in the MC phase this fact entails that the maximal current ${J = 1/4}$ is actually reached only at the right boundary (at odds with the pure TASEP, in which the current value is spatially invariant). 
Upon decreasing the extraction rate to ${\beta < 1/2}$, the MC phase evolves into the HD phase through a continuous transition (denoted by a dashed line in figure~\ref{fig:phase_diagram}), which is still quite similar to that of the pure TASEP, even though here it can also extend to ${\alpha < 1/2}$. 
The most remarkable effect of Langmuir kinetics is however the onset of parameter regions where two of the pure phases (namely LD/HD or LD/MC in the present case ${\Omega_\mathrm{A} > \Omega_\mathrm{D}}$) coexist, being separated by a static domain wall, i.e.~a domain wall remaining localized and stable in the steady state~\cite{ParmeggianiFranoschFrey03,ParmeggianiFranoschFrey04}. 
Upon increasing the injection rate $\alpha$, such domain wall moves towards the left boundary of the system and, when it reaches the boundary, coexistence terminates and the system falls into a pure HD or MC steady state, depending on the $\beta$ value. 
In figure~\ref{fig:phase_diagram}, the separation between the LD/HD and the LD/MC coexistence regions is denoted by a dotted line, whereas the separation between the coexistence regions and the pure HD and MC phase regions is denoted by a solid line. 
On the other hand, upon decreasing $\alpha$, the domain wall moves towards the right boundary of the system and, when it reaches the boundary, the steady state falls into a pure LD phase. 
Let us note that the separation between the LD/MC coexistence and the pure LD phase region, denoted in figure~\ref{fig:phase_diagram} by a dash-dotted line, exhibits indeed a special feature, that is a domain-wall amplitude vanishing exactly while reaching the (right) boundary of the system, i.e.~where coexistence terminates. 
It has also been observed that, moving along the LD phase boundary on increasing $\beta$, the domain-wall amplitude vanishes \emph{continuously} at ${\beta = 1/2}$, so that in \cite{ParmeggianiFranoschFrey04} this point of the phase diagram has been likened to an equilibrium critical point. 
Let us finally recall that the phase diagram that we have described corresponds to moderate values of the attachment rate $\Omega_\mathrm{A}$; increasing $\Omega_\mathrm{A}$ beyond a certain threshold leads to a disappearance of the pure LD phase.

\begin{figure}
	\centering
	\resizebox{106mm}{!}{\includegraphics*{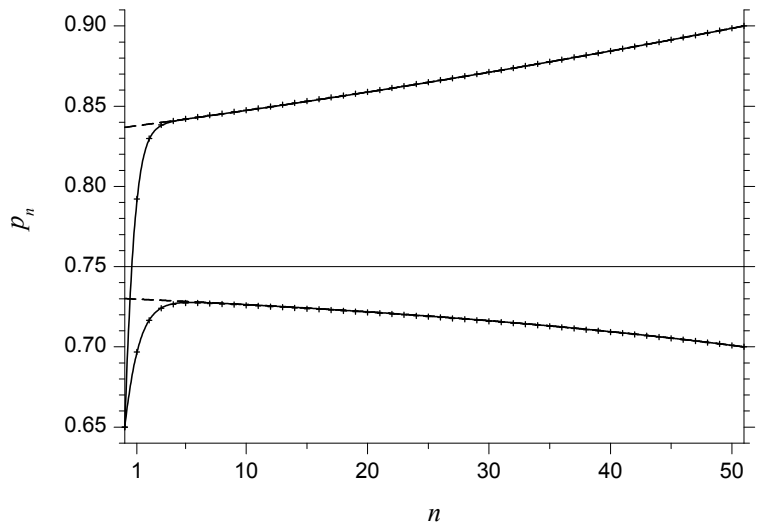}}
	\caption
	{
		Steady-state local density $\dfull_n$ as a function of the node index $n$ for a system size ${N = 50}$ and the following parameter values: ${\Omega_\mathrm{D} = 0.1}$, ${\Omega_\mathrm{A} = 0.3}$, ${\alpha = 0.65}$, ${\beta = 0.1}$ (top graph) and ${\beta = 0.3}$ (bottom graph). 
		Symbols denote numerical results, whereas solid and dashed lines denote respectively the full analytical expression and the bulk term alone (see the text). 
		A (horizontal) thin solid line denotes the equilibrium Langmuir density \eqref{eq:Langmuir_isotherm}.
	}
	\label{fig:density_profiles}
\end{figure}

The steady-state density profiles in the different phases have been described in \cite{ParmeggianiFranoschFrey04} in full detail. 
Here we briefly recall what happens in the HD phase, which we are mainly interested in. 
Figure~\ref{fig:density_profiles} displays the density profiles for two points in the HD phase. 
Note that such profiles can be computed by two different methods, namely either solving numerically the finite-size mean-field equations \eqref{eq:p_steady_state} (with the appropriate boundary conditions \eqref{eq:p_boundaries}) or by an analytical expression, discussed in~\ref{sec:appendix_dens_prof}, which holds asymptotically for large $N$. 
By the way, figure~\ref{fig:density_profiles} clearly shows that the latter is quantitatively very accurate, even for quite small system sizes, such as ${N = 50}$. 
The analytical expression is in principle quite similar to the one obtained for the balanced case~\cite{BPPZ18}, in that it is made up of a sum of a bulk term plus a boundary-layer term. 
In practice, the bulk term is the solution of the differential equation arising in the continuum (hydrodynamic) limit~\cite{ParmeggianiFranoschFrey04}, whereas the boundary-layer term is the mean-field solution for an ``effective'' pure TASEP, where ``effective'' means that the uniform current of the pure TASEP is replaced by a local current value~\cite{BPPZ18}. 
As apparent from figure~\ref{fig:density_profiles}, it turns out that the bulk profile is always \emph{all above} or \emph{all below} the equilibrium Langmuir density 
\begin{equation}
\ell = \frac{\Omega_\mathrm{A}}{\Omega_\mathrm{A} + \Omega_\mathrm{D}}
\, ,
\label{eq:Langmuir_isotherm}
\end{equation}
depending on whether $\oneminus{\beta}$ (the right-boundary density) is respectively larger or smaller than $\ell$ (when ${\oneminus{\beta} = \ell}$ the profile is flat and coinciding with the Langmuir density). 
According to the theory~\cite{ParmeggianiFranoschFrey04}, the corresponding left-boundary density of the bulk profile (i.e.~the intercept of the dashed lines with the vertical axis ${n=0}$), which we shall denote by $\vartheta$, can be written as a function of $\beta$, $\Omega_\mathrm{A}$ and $\Omega_\mathrm{D}$ (not $\alpha$) as 
\begin{equation}
\frac{\vartheta - \ell  \vphantom{\frac{1}{2}}}{\ell - \frac{1}{2}}
= 
{W}_{0} \left(
\frac{\oneminus{\beta} - \ell  \vphantom{\frac{1}{2}}}{\ell - \frac{1}{2}}
\, \exp \frac{\oneminus{\beta} - \ell - \Omegamedio}{\ell - \frac{1}{2}} 
\right)
\, ,
\label{eq:definizione_theta}
\end{equation}
where $W_{0}$ is a Lambert function (precisely, the so-called \emph{zero branch}).
More generally, the Lambert function allows one to express the whole bulk profile~\cite{ParmeggianiFranoschFrey04}, showing in particular that the profile slope is always \emph{everywhere positive} or \emph{everywhere negative}, respectively in the two aforementioned cases. 
Of course, boundary layers may locally reverse the density profile slope, as one can actually observe in figure~\ref{fig:density_profiles} in the bottom graph. 
Let us finally remark the fact that the \emph{bulk profile} is not only a mathematical concept, but it also has an important physical role. 
Indeed it reflects the behavior of the current (which conversely is not affected by boundary layers) and therefore it is relevant for phase transitions, including dynamical ones, as we shall see in the next section.

\section{The dynamical transitions}
\label{sec:dyn_trans}

In this section we present the main original results of our work, dealing in particular with the spectrum of the mean-field relaxation matrix and with the large-$N$ asymptotic behavior of its smallest eigenvalue (slowest relaxation rate). 
We concentrate on the HD phase because, as previously mentioned, it is the one where more interesting effects can be observed. 
The LD phase behavior will be briefly summarized at the end of this section. 
Let us recall, however, that all the results still refer to a case ${\Omega_\mathrm{A} > \Omega_\mathrm{D}}$, here in particular ${\Omega_\mathrm{D} = 0.1}$ and ${\Omega_\mathrm{A} = 0.3}$, and that the roles of HD and LD would be exchanged, if we took the opposite assumption ${\Omega_\mathrm{A} < \Omega_\mathrm{D}}$. 
The latter case does not need to be investigated, because of the particle-hole symmetry mentioned above.
Let us also recall that, according to \cite{ParmeggianiFranoschFrey04}, for ${\Omega_\mathrm{A} > \Omega_\mathrm{D}}$ the HD phase corresponds to the parameter region defined by the following inequalities
\begin{subequations} 
\label{eq:intervalli_valori_HD}
	\begin{eqnarray}
	\textstyle \frac{1}{2} > \beta > 0 
	\, , \label{eq:intervallo_valori_beta} \\
	\alpha > \oneminus{\vartheta} 
	\, , \label{eq:intervallo_valori_alpha_vs_beta}
	\end{eqnarray}
\end{subequations} 
where $\vartheta$ depends on $\beta$ (besides $\Omega_\mathrm{A}$ and $\Omega_\mathrm{D}$) according to \eqref{eq:definizione_theta}.

\subsection{Spectrum of the relaxation matrix}
\label{subsec:spectrum}

Let us first describe the results that we have obtained by solving the eigenvalue problem \eqref{eq:eigenvalue_equation_symmetrized} numerically at finite $N$.
As previously mentioned, it turns out that all eigenvalues are real, as we deal with a real symmetric matrix, which has been proved to be similar to the mean-field relaxation matrix. 
The most relevant novelty item with respect to the balanced case is that the low-lying part of the spectrum turns out to behave in two qualitatively different manners, depending on whether the extraction rate $\beta$ is smaller or larger than $\oneminus{\ell}$, where $\ell$ is the equilibrium Langmuir density. 
As discussed in the previous section, these two regimes are respectively associated to increasing or decreasing bulk density profiles, and therefore (see \ref{sec:appendix_dens_prof}) \emph{decreasing} or \emph{increasing} current profiles. 
Note in particular that the latter type of behaviour never occurs in the balanced case. 

\begin{figure}
	\centering
	\resizebox{106mm}{!}{\includegraphics*{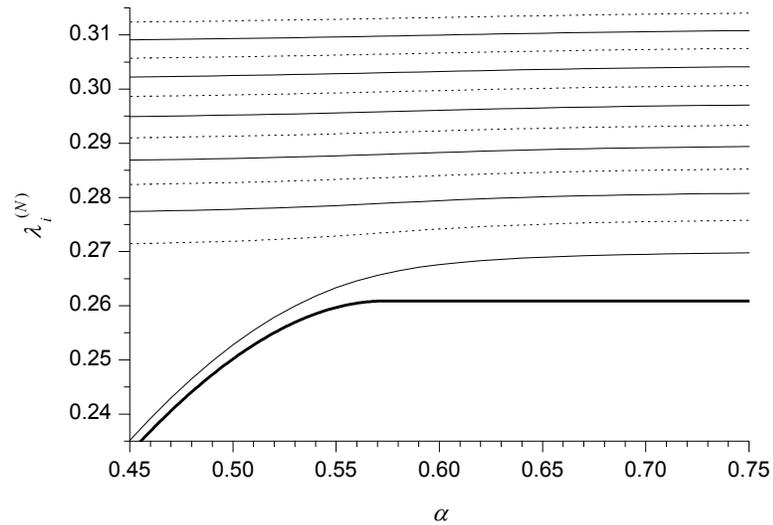}}
	\caption
	{
		Thin lines represent the smallest $12$ eigenvalues of the relaxation matrix, computed numerically for ${N = 300}$ and ${\beta = 0.1}$, as a function of $\alpha$. 
		Eigenvalues of increasing magnitude are alternately displayed by solid and dotted lines. 
		The thick solid line represents the ${N \to \infty}$ limit of the smallest eigenvalue, i.e.~the function defined by \eqref{eq:slowest_rate}, which in this case coincides with \eqref{eq:slowest_rate_second} (see the text).
	}
	\label{fig:autovalori_beta=0-1}
\end{figure}

For ${\beta < \oneminus{\ell}}$ the situation, as a function of the injection rate $\alpha$, is that displayed in figure~\ref{fig:autovalori_beta=0-1}. 
We can see that the eigenvalues are arranged in a nearly ``flat band'', i.e.~they are all rather weakly dependent on $\alpha$, with the exception of the smallest eigenvalue. 
This one exhibits indeed a clearly increasing trend upon increasing $\alpha$ up to a certain critical region, beyond which it saturates, almost becoming constant. 
In this case the mean-field spectrum is qualitatively similar to that observed in the balanced case~\cite{BPPZ18} and in the pure TASEP~\cite{PelizzolaPretti17} as well. 
In analogy with such cases, we expect that the infinite-size limit of the smallest eigenvalue becomes actually constant beyond a critical $\alpha$ value, the latter being precisely characterized by a discontinuity in the second derivative. 

\begin{figure}
	\centering
	\resizebox{106mm}{!}{\includegraphics*{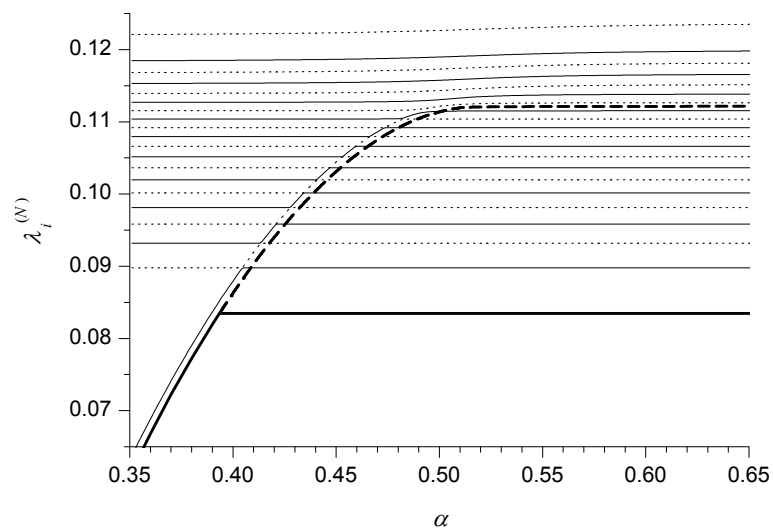}}
	\caption
	{
		Thin lines represent the smallest $20$ eigenvalues of the relaxation matrix, computed numerically for ${N = 300}$ and ${\beta = 0.3}$, as a function of $\alpha$. 
		Eigenvalues of increasing magnitude are alternately displayed by solid and dotted lines. 
		The thick solid line represents the ${N \to \infty}$ limit of the smallest eigenvalue, i.e.~the function defined by \eqref{eq:slowest_rate}.
		The thick dashed line represents \eqref{eq:slowest_rate_second}. 
	}
	\label{fig:autovalori_beta=0-3}
\end{figure}

Figure~\ref{fig:autovalori_beta=0-3} displays the typical scenario for ${\beta > \oneminus{\ell}}$. 
At first glance, this case could be roughly described by saying that the isolated eigenvalue (being also the smallest one in the low-$\alpha$ region) does not ``join the band'' smoothly, but rather it ``enters the band'', giving rise to a number of degeneracies or \emph{level crossings}. 
Nevertheless, by examining the results on an appropriately small scale, one realizes that in fact there is no degeneracy at all (at finite $N$), that is, the seeming crossings are in fact all \emph{avoided crossings}. 
As a consequence, the smallest eigenvalue undergoes in fact an abrupt slope change as a function of $\alpha$, whereas a number of higher eigenvalues exhibit \emph{two} subsequent slope changes. 
As we shall see below, in the ${N \to \infty}$ limit the abrupt slope change of the smallest eigenvalue becomes an actual discontinuity of the first derivative.

\subsection{Analytical results}
\label{subsec:analytics}

Let us now present the analytical results. 
Note that in this section we only discuss the final results and their physical meaning, whereas all technical details of the derivation are reported in \ref{sec:appendix_eigenvalues}. 
Let us denote by ${\lambda_{\min}}^{(N)}$ the slowest relaxation rate for a system of size $N$ (from now on we indicate the size dependence by a superscript), and let us define the infinite-size limit 
\begin{equation}
{\lambda_{\min}}^{(\infty)} 
\equiv \lim_{N \to \infty} {\lambda_{\min}}^{(N)} 
\, .
\label{eq:slowest_rate_limit}
\end{equation}
The central result is that the latter quantity can be written as
\begin{equation}
{\lambda_{\min}}^{(\infty)} 
= 1 - 2 \max \left\{ 
{x}_{*} \sqrt{ \vphantom{\beta} \vartheta \oneminus{\vartheta} } , 
\sqrt{ \beta \oneminus{\beta} } 
\, \right\}
\, ,
\label{eq:slowest_rate}
\end{equation}
where ${\vartheta = \vartheta(\beta,\Omega_\mathrm{A},\Omega_\mathrm{D})}$ is the left-boundary value of the bulk density, determined by \eqref{eq:definizione_theta}, and ${{x}_{*} = {x}_{*}(\alpha,\vartheta)}$ is determined by the behavior of a real function $f(x;\alpha,\vartheta)$ of the real variable ${x \ge 1}$ and of the model parameters $\alpha$, $\beta$, $\Omega_\mathrm{A}$ and $\Omega_\mathrm{D}$ (the last three contained in $\vartheta$). 
In full analogy with the balanced case~\cite{BPPZ18}, the $f$ function can be defined as
\begin{equation}
f(x) \equiv \sum_{n=1}^{\infty} 
\frac{ \ddlim_{n+1} - \ddlim_{n-1} }{ \sqrt{ \vartheta \oneminus{\vartheta} } }
\, {v}_{n}(x) {\zeta(x)}^{n}
\, , 
\label{eq:definizione_fx_come_serie}
\end{equation}
where
\begin{equation}
\zeta(x) \equiv x - \sqrt{x^2-1}
\, , 
\label{eq:definizione_z}
\end{equation}
while the sequences $\ddlim_{n}$ and ${v}_{n}(x)$ (depending on the model parameters) are defined by recursion, respectively as
\begin{equation}
\ddlim_{0} \equiv \alpha 
\, , \qquad 
\ddlim_{n+1} \equiv 1 - \frac{\vartheta \oneminus{\vartheta}}{\ddlim_{n}} 
\qquad n = 0,1,2,\dots
\, , 
\label{eq:definizione_s_ricorsiva}
\end{equation}
and
\begin{eqnarray}
{v}_{0}(x) \equiv 0 \, , \qquad {v}_{1}(x) \equiv 1 
\, , \label{eq:definizione_v} \\
{v}_{n+1}(x) \equiv \left( 2x 
- \frac{ \ddlim_{n+1} - \ddlim_{n-1} }{ \sqrt{ \vartheta \oneminus{\vartheta} } } 
\right) 
{v}_{n}(x) - {v}_{n-1}(x)
\qquad n = 1,2,\dots
\, . \nonumber
\end{eqnarray}
The series in \eqref{eq:definizione_fx_come_serie} turns out to converge very quickly, which makes it amenable to numerical evaluation with extremely high precision, and at a negligible computational cost. 
Without entering the details, let us note that the model parameters $\beta$, $\Omega_\mathrm{A}$ and $\Omega_\mathrm{D}$ get into the theory only through the quantities ${\beta \oneminus{\beta}}$ and ${\vartheta \oneminus{\vartheta}}$, representing respectively the right- and left-boundary currents in the infinite-size limit (see \ref{sec:appendix_dens_prof}). 
The sequence $\ddlim_{n}$ represents the mean-field density profile for a pure TASEP with current ${\vartheta \oneminus{\vartheta}}$ (still in the infinite-size limit), and indeed the recursion equation \eqref{eq:definizione_s_ricorsiva} coincides with the one reported in the early work by Derrida, Domany and Mukamel~\cite{DerridaDomanyMukamel92}. 
This is related to the fact (mentioned in the previous section and explained in \ref{sec:appendix_dens_prof}) that in the TASEP-LK the boundary layer behaves just like in the pure TASEP, yet with a \emph{local} current value. 
Let us also observe that, from a formal point of view, equations \eqref{eq:definizione_fx_come_serie}--\eqref{eq:definizione_v} are really \emph{identical} to those of the balanced case~\cite{BPPZ18} (which in turn also apply to the pure TASEP). 
The difference lies only in the parameter $\vartheta$ (i.e.~the left-boundary bulk density), whose expressions for the balanced case and for the pure TASEP were respectively $\oneminus{(\beta + \Omega)}$ and $\oneminus{\beta}$.
The behavior of ${x}_{*}$ turns out to be consequently analogous, as we outline below (for more details see \cite{BPPZ18}).\footnote
{The analogy is in fact very precise, as it consists in making the substitution ${\beta \to \beta+\Omega}$ (from the pure TASEP to the balanced TASEP-LK) or ${\beta \to \oneminus{\vartheta}}$ (from the pure TASEP to the unbalanced TASEP-LK).}

As previously mentioned, we assume that $\Omega_\mathrm{A}$ and $\Omega_\mathrm{D}$ are fixed, so that $\vartheta$ only depends on $\beta$, according to \eqref{eq:definizione_theta}. 
Given a particular value of $\beta$, there turns out to exist an interval of $\alpha$ values, larger than a critical threshold $\alpha_\mathrm{c}$, such that 
\begin{equation}
f(x;\alpha,\vartheta) < 1
\qquad \forall x \geq 1
\label{eq:f_disuguaglianza}
\, .
\end{equation}
In all this interval we have ${x_* = 1}$, independently of $\alpha$. 
Otherwise, when $\alpha$ becomes smaller than $\alpha_\mathrm{c}$, condition \eqref{eq:f_disuguaglianza} no longer holds, and in particular ${f(1;\alpha,\vartheta) > 1}$. 
In this case we have ${x_* > 1}$, and the precise value of $x_*$ is determined by the equation 
\begin{equation}
f(x_*;\alpha,\vartheta) = 1
\label{eq:f_equazione}
\, .
\end{equation}
Still at fixed $\beta$ (i.e.~fixed $\vartheta$), the $x_*$ value depends on $\alpha$, in particular $x_*$ decreases upon increasing $\alpha$. 
The critical threshold $\alpha_\mathrm{c}(\vartheta)$ is determined by the equation
\begin{equation}
f(1;\alpha_\mathrm{c},\vartheta) = 1
\label{eq:f_equazione_soglia}
\, .
\end{equation}
In the end $x_*$ turns out to be (at fixed $\vartheta$) a continuous function of $\alpha$, with a discontinuity in the second derivative (second-order singularity) at ${\alpha = \alpha_\mathrm{c}}$. 

Let us now consider equation \eqref{eq:slowest_rate}. 
As discussed in the previous section, in the low-$\beta$ region (${\beta < \oneminus{\ell}}$) the bulk density profile has a positive slope, and hence ${\oneminus{\beta} > \vartheta}$. 
Since we are in the HD phase, both $\oneminus{\beta}$ and $\vartheta$ are larger than $1/2$, which entails ${\vartheta \oneminus{\vartheta} > \beta \oneminus{\beta}}$. 
Now, as we have previously seen that ${{x}_{*} \ge 1}$, it obviously follows that \eqref{eq:slowest_rate} simplifies to
\begin{equation}
{\lambda_{\min}}^{(\infty)} 
= 1 - 2 {x}_{*} \sqrt{ \vphantom{\beta} \vartheta \oneminus{\vartheta} }
\, .
\label{eq:slowest_rate_second}
\end{equation}
In this regime, the behaviour of ${\lambda_{\min}}^{(\infty)}$ as a function of $\alpha$ is only governed by ${x}_{*}$ (which gives rise to the aforementioned second-order singularity), and therefore it turns out to be qualitatively similar to that of the balanced case, as displayed in figure~\ref{fig:autovalori_beta=0-1}. 
Conversely, in the high-$\beta$ region (${\beta > \oneminus{\ell}}$), the bulk density profile has a negative slope, so that all inequalities are reversed, yielding ${\vartheta \oneminus{\vartheta} < \beta \oneminus{\beta}}$. 
Since we have previously seen that ${x}_{*}$ decreases upon increasing $\alpha$ for ${\alpha < \alpha_\mathrm{c}}$ and remains constantly equal to $1$ for ${\alpha \ge \alpha_\mathrm{c}}$, we can argue that in this case there necessarily exists another threshold value ${\tilde{\alpha}_\mathrm{c} < \alpha_\mathrm{c}}$, such that ${\sqrt{\beta \oneminus{\beta}} > {x}_{*} \sqrt{\vartheta \oneminus{\vartheta}}}$ for ${\alpha > \tilde{\alpha}_\mathrm{c}}$. 
This latter threshold is of course determined by the equation
\begin{equation}
f\left(\!\textstyle \sqrt{\frac{\beta \oneminus{\beta}}{\vartheta \oneminus{\vartheta}}};
\tilde{\alpha}_\mathrm{c},\vartheta\right) = 1
\label{eq:f_equazione_soglia_tilde}
\, ,
\end{equation}
from which one can also argue that $\tilde{\alpha}_\mathrm{c}(\vartheta,\beta)$ shall coincide with ${\alpha}_\mathrm{c}(\vartheta)$ when ${\beta = \oneminus{\ell} = \oneminus{\vartheta}}$ (that is when the bulk density profile is flat and coinciding with the equilibrium Langmuir density). 
In this regime, which has no analogue in the balanced case, ${\lambda_{\min}}^{(\infty)}$ as a function of $\alpha$ develops a discontinuity in the first derivative (first-order singularity) at ${\alpha = \tilde{\alpha}_\mathrm{c}}$, as displayed in figure~\ref{fig:autovalori_beta=0-3}. 
We also notice an intriguing fact, namely that for ${\alpha > \tilde{\alpha}_\mathrm{c}}$ the analytical expression \eqref{eq:slowest_rate_second}, while moving away from the smallest eigenvalue, closely follows the whole series of avoided crossings. 
Moreover, the critical value ${\alpha}_\mathrm{c}$, where \eqref{eq:slowest_rate_second} exhibits the second-order singularity, can be roughly identified with the end of the crossing regime. 

\begin{figure}
	\centering
	\resizebox{106mm}{!}{\includegraphics*{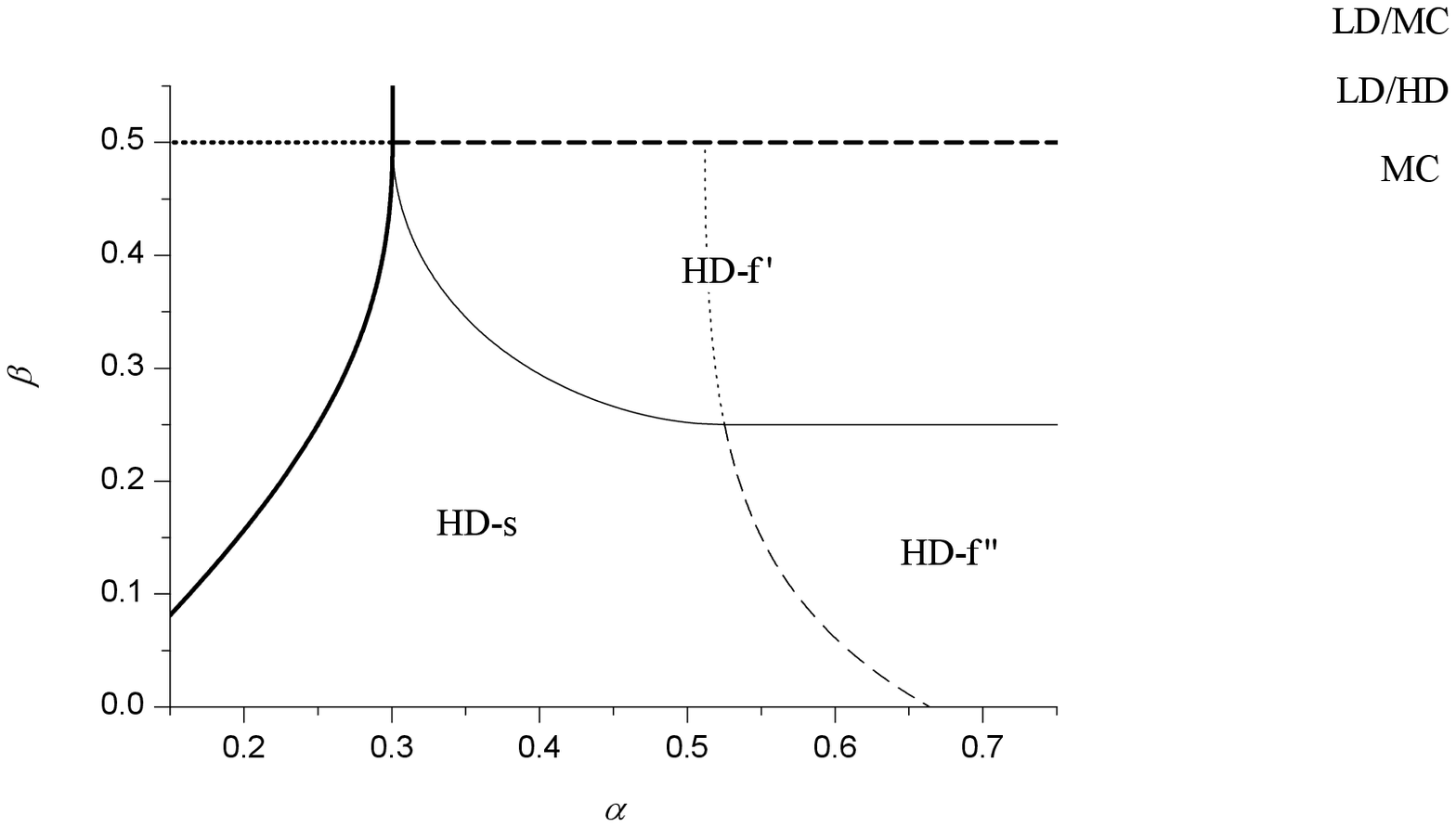}}
	\caption
	{
		Phase diagram in the $\alpha$-$\beta$ plane (HD phase region) for ${\Omega_\mathrm{D} = 0.1}$ and ${\Omega_\mathrm{A} = 0.3}$. 
		Thick lines denote static transitions (line types as in figure~\ref{fig:phase_diagram}), thin lines denote dynamical transitions (line types and subphase labels explained in the text). 
	}
	\label{fig:phase_diagram_HD}
\end{figure}

We summarize all the analytical results in a \emph{dynamical} phase diagram (figure \ref{fig:phase_diagram_HD}), partitioning the HD phase into regions (subphases), whose border lines coincide with the singularities described above. 
We first distinguish two main subphases, denoted as HD-s (``slow'') and HD-f (``fast''), which respectively correspond to the regions where ${\lambda_{\min}}^{(\infty)}$ depends on $\alpha$ or not (the terms ``slow'' and ``fast'' to denote these subphases have been previously used in \cite{BottoPelizzolaPretti18,PelizzolaPrettiPuccioni19}). 
In particular, in the fast phase ${\lambda_{\min}}^{(\infty)}$ is not only independent of $\alpha$ but it also takes the highest admissible value at given $\beta$, namely  
\begin{equation}
{\lambda_{\min}}^{(\infty)} 
= 1 - 2 \max \left\{ 
\! \sqrt{ \vphantom{\beta} \vartheta \oneminus{\vartheta} } , 
\sqrt{ \beta \oneminus{\beta} } 
\, \right\}
\, .
\label{eq:slowest_rate_first}
\end{equation}
This last expression exhibits a first-order singularity at ${\beta = \oneminus{\ell} = \oneminus{\vartheta}}$ (${= 0.25}$ in figure \ref{fig:phase_diagram_HD}), which justifies a further subdivision of the HD-f region into HD-f$\,'$ and HD-f$\,''$, or in other words the existence of one more dynamical transition.
Note that in this case, as the right-hand side of \eqref{eq:slowest_rate_first} no longer depends on $\alpha$, the singularity is indeed a discontinuity in the first derivative with respect to $\beta$. 
In figure \ref{fig:phase_diagram_HD} we denote dynamical transitions, characterized by first- or second-order singularities, respectively by solid or dashed (thin) lines. 
According to the above discussion, the transition between HD-s and HD-f$\,'$ is of the former type, and occurs at ${\alpha = \tilde{\alpha}_\mathrm{c}}$, whereas the transition between HD-s and HD-f$\,''$ is of the latter type, and occurs on the line ${\alpha = {\alpha}_\mathrm{c}}$. 
Concerning this last curve, let us finally note that in figure \ref{fig:phase_diagram_HD} we have also displayed (as a dotted line) its continuation for ${\beta > \oneminus{\ell}}$. 
As previously mentioned, this section of the line cannot be properly said to represent a dynamical transition, because it is unrelated to the slowest relaxation rate, yet it contains some extra information about the spectrum, in that it can be loosely considered as a border line for the level-crossing region.

\begin{figure}
	\centering
	\resizebox{106mm}{!}{\includegraphics*{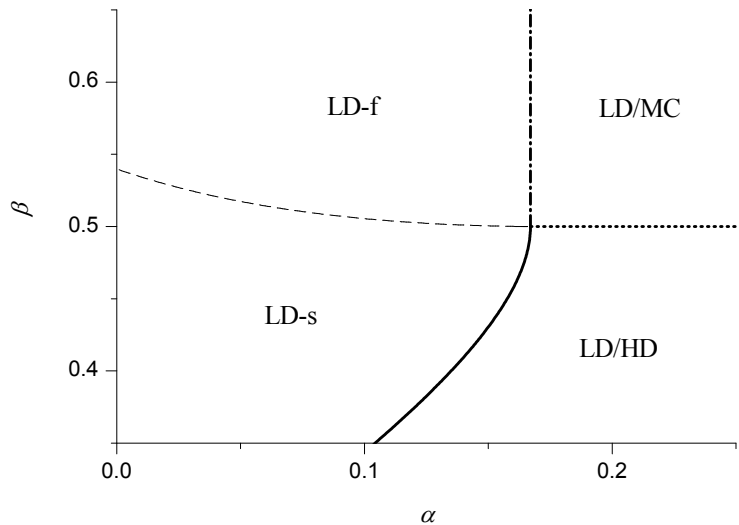}}
	\caption
	{
		Phase diagram in the $\alpha$-$\beta$ plane (LD phase region) for ${\Omega_\mathrm{D} = 0.1}$ and ${\Omega_\mathrm{A} = 0.2}$. 
		Thick lines denote static transitions (line types as in figure~\ref{fig:phase_diagram}), a thin dashed line denotes the (second-order-like) dynamical transition (subphase labels explained in the text). 
	}
	\label{fig:phase_diagram_LD}
\end{figure}

To conclude this subsection, we give a brief account of the dynamical transition scenario also in the LD phase (figure~\ref{fig:phase_diagram_LD}).\footnote{Let us recall, once again, that these results refer to a case ${\Omega_\mathrm{A} > \Omega_\mathrm{D}}$, and that the roles of HD and LD would be exchanged, if we took the opposite assumption ${\Omega_\mathrm{A} < \Omega_\mathrm{D}}$.} 
We do not report any analytical detail about this case, which is indeed conceptually analogous, the basic differences being only that the bulk density profile is described by the other real branch of the Lambert function~\cite{ParmeggianiFranoschFrey04} and that the boundary layer takes place at the right, rather than left, boundary. 
In the LD phase the bulk density profile is always a monotonically increasing function of the node index $n$, and this property gives rise to a much simpler transition scenario, with a unique transition line characterized by a second-order singularity. 
Such a transition divides the LD phase into slow and fast subphases (labelled as {LD-s} and {LD-f}), respectively characterized by the property that ${\lambda_{\min}}^{(\infty)}$ depends on $\beta$ or not. 
Accordingly, the singularity appears as a discontinuity in the (second) derivative with respect to $\beta$. 
A noticeable fact is that, at the LD-phase boundary, the dynamical transition line terminates precisely at the pseudo-critical point, which we mentioned in section~\ref{sec:phase}, where the amplitude of the static domain wall vanishes.
We do not have, however, an intuitive explanation of this fact.

\subsection{Other numerical results}

In this subsection we deal with some more issues that may be of interest for the system under investigation, but which go beyond the analytical treatment presented above. 

\begin{figure}
	\centering
	\resizebox{106mm}{!}{\includegraphics*{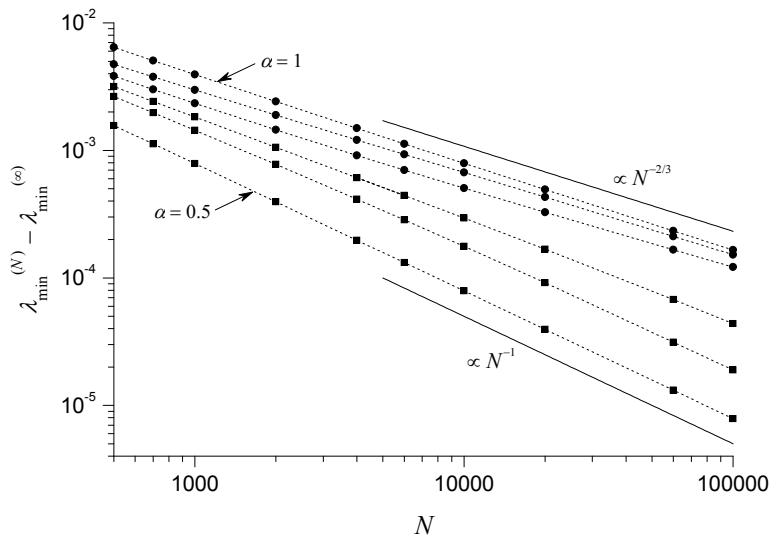}}
	\caption
	{
		Difference between the finite-size slowest relaxation rate ${\lambda_{\min}}^{(N)}$ and its infinite-size limit ${\lambda_{\min}}^{(\infty)}$ as a function of $N$ for ${\Omega_\mathrm{D} = 0.1}$, ${\Omega_\mathrm{A} = 0.3}$, ${\beta = 0.1}$ and different $\alpha$ values ($\alpha = 0.5, 0.56, 0.57, 0.58, 0.6, 1$), below (squares) and above (circles) the critical threshold ${\alpha_\mathrm{c} \approx 0.573676}$. 
		The solid lines represent scaling functions. 
		The dotted lines are a guide for the eye.
	}
	\label{fig:scaling_beta_01}
\end{figure}

In the first place, we consider the asymptotic scaling behavior of the slowest relaxation rate at large $N$, specifically in order to verify whether the scaling exponents are the same as those obtained in the balanced case. 
A comparison between the numerical results and the trial scaling functions is reported in figure \ref{fig:scaling_beta_01} for the low-$\beta$ regime (${\beta < \oneminus{\ell}}$), where we recall that the dynamical transition is second-order-like. 
The results are apparently well compatible with a scenario in which the quantity ${{\lambda_{\min}}^{(N)} - {\lambda_{\min}}^{(\infty)}}$ scales with the inverse of $N$ as a power law with the same balanced-case exponents, namely $1$ in the HD-s phase ($\alpha < \alpha_\mathrm{c}$) and $2/3$ in the HD-f$\,''$ phase ($\alpha > \alpha_\mathrm{c}$). 
We have performed a similar analysis in the high-$\beta$ regime (${\beta > \oneminus{\ell}}$) and the results (not shown) clearly indicate that even the HD-f$\,'$ phase is characterized by power-law scaling with exponent $2/3$.  
Let us recall that, as far as the pure TASEP is concerned, the exact result \cite{deGierEssler05} predicts a power-law behavior with a unique scaling exponent $2$, being unaffected by the dynamical transition, whereas the mean-field theory predicts the same exponent, though (curiously) only in the \emph{fast} phase. 
The current paper, along with \cite{BPPZ18}, provide considerable evidence that the mean-field scaling exponent is $2/3$ in the HD-f (and LD-f as well) of both the balanced and unbalanced TASEP-LK. 
All these results together lead us to conjecture that $2/3$ might indeed be the exact scaling exponent for the TASEP-LK in the whole HD and LD phases.

\begin{figure}
	\centering
	\resizebox{106mm}{!}{\includegraphics*{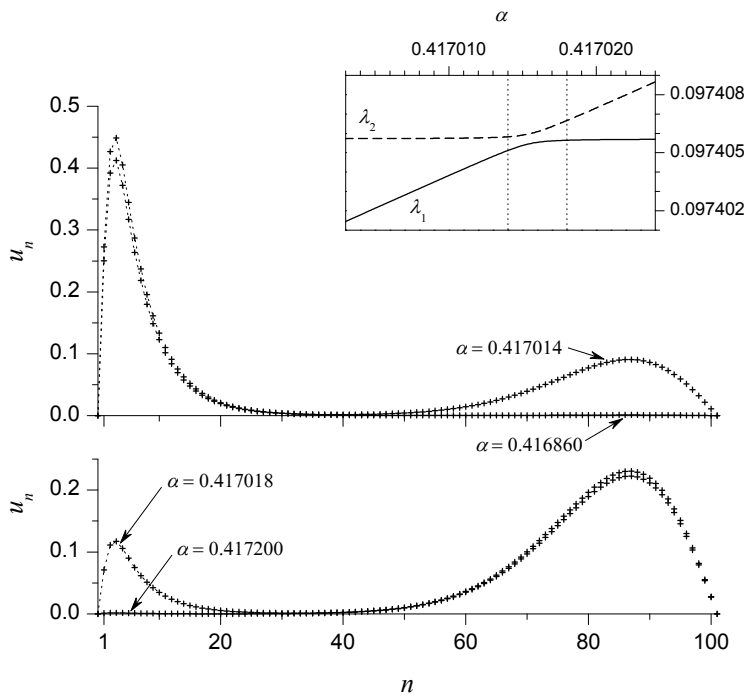}}
	\caption
	{
		The symbols (crosses) denote the components $u_n$ of the (normalized) eigenvector corresponding to the smallest eigenvalue for ${N = 100}$, ${\Omega_\mathrm{D} = 0.1}$, ${\Omega_\mathrm{A} = 0.3}$, ${\beta = 0.3}$ and different $\alpha$ values, close to the (first-order-like) dynamical transition. 
		The dotted lines are a guide for the eye.
		The inset displays the smallest ($\lambda_1$, solid line) and second smallest ($\lambda_2$, dashed line) eigenvalues as a function of $\alpha$. 
	}
	\label{fig:autovettori_beta=0-3}
\end{figure}

Another interesting point is that, according to our mean-field theory, the first-order-like dynamical transition turns out to be accompanied by a structural change in the relaxation mode associated to the slowest relaxation rate (or, in mathematical terms, the eigenvector associated to the smallest eigenvalue of the relaxation matrix). 
This is not the case in an ordinary (i.e.~second-order-like) dynamical transition. 
To get an example of this behavior, we solve the eigenvalue problem for a system of finite size $N$, fixing $\beta$ (in the high-$\beta$ regime of the HD phase) and varying $\alpha$. 
The results are reported in figure~\ref{fig:autovettori_beta=0-3}. 
We can identify a transition region, around ${\alpha \approx 0.417016}$, where the eigenvector exhibits a bimodal shape, being characterized by two different ``peaks'' of (conventionally) positive components at both low and high $n$ values. 
The bimodality region clearly corresponds to the avoided crossing region, where the first and second smallest eigenvalues are nearly degenerate (see the inset in figure~\ref{fig:autovettori_beta=0-3}). 
Note that, due to the finite size, the bimodality region has a finite extension in $\alpha$ and it is not so close to the theoretical (infinite-size) threshold ${\tilde{\alpha}_\mathrm{c} \approx 0.393451}$. 
However, upon increasing $N$, one can verify numerically that the transition region tends to collapse, and it does so precisely towards $\tilde{\alpha}_\mathrm{c}$. 
Out of this region, the eigenvector rapidly becomes unimodal, and in particular the low-$n$ peak survives at low $\alpha$, whereas the high-$n$ peak survives at high $\alpha$. 
Moreover, the eigenvector associated to the second smallest eigenvalue exhibits a similar but opposite behavior: it is still bimodal in the transition region\footnote{For orthogonality reasons, in this case the eigenvector components do not have a definite sign.} and unimodal elsewhere, but the low-$n$ peak survives at \emph{high} $\alpha$, whereas the high-$n$ peak survives at \emph{low} $\alpha$.  
In other words, one could state that the two relaxation modes exchange their respective structure concurrently with the avoided crossing. 
This mechanism, which similarly takes place at every subsequent avoided crossing, can provide a qualitative explanation of a previously observed fact, namely that the analytical expression \eqref{eq:slowest_rate_second} closely follows the sequence of avoided crossings. 
The reason is indeed that this expression, ultimately based on approximating the eigenvector with the sequence $v_n(x_*)$, describes a specific eigenmode structure, characterized by a low-$n$ peak. 
According to the above discussion, one can argue that such a structure does not remain associated to the smallest eigenvalue (as it happens in the second-order-like scenario), but it is progressively ``transferred'' to higher eigenvalues. 

Let us make a brief comment about these last results, in particular about the possibility of observing dynamical transitions in real (or simulated) kinetics. 
It is a known fact that the ordinary (second-order-like) dynamical transition occurring in the pure TASEP, whose existence has been proved rigorously~\cite{deGierEssler06}, is nonetheless difficult to detect in simulations~\cite{ProemeBlytheEvans11}, because the effect is weak and it is easily overcome by noise. 
Based on the mean-field theory, in \cite{BPPZ18} we suggested that such a difficulty could also be related to the fact that, concurrently with the dynamical transition, the slowest relaxation mode joins the band of higher-order modes, whose level spacing vanishes in the thermodynamic limit. 
As a consequence, entering the fast phase, simulations inevitably begin to observe a superposition of modes, resulting in a smoothed transition. 
In principle, one could expect that the eigenmode structure change associated with the first-order-like dynamical transition could make it more clearly detectable. 
Unfortunately, we must remember the results presented above refer to the eigenvectors of the symmetrized relaxation matrix, and that in order to obtain those of the original matrix (i.e.~the physically relevant ones), we have to invert the similarity transformation \eqref{eq:similarity_transformation}. 
Qualitatively speaking, the latter operation amounts to superimposing an exponential decay (with respect to $n$), which would result in a nearly complete suppression of the right-boundary peak. 
In conclusion we expect that the first-order-like dynamical transition might present the same detection difficulties as the second-order one. 


\section{Summary and conclusions}
\label{sec:conclusion}

We have investigated dynamical transitions in the open TASEP with unbalanced Langmuir kinetics. 
This is a generalization of the TASEP, such that particles can also bind to empty nodes or unbind from occupied ones. 
When the rates of these additional processes (which we denote as binding and unbinding rates) are equal, we say that the Langmuir
kinetics is balanced, otherwise it is called unbalanced. 
The model cannot be solved exactly, but its stationary state is well known from~\cite{ParmeggianiFranoschFrey03,Popkov03,Evans03,ParmeggianiFranoschFrey04}. 
In this paper we focused on purely dynamical transitions, that is on singularities of the relaxation rate that are not associated to any stationary-state transition. 
Such dynamical transitions occur in the high- and low-density phases, splitting each of them into a \emph{slow} subphase, where the relaxation rate depends on both boundary reservoirs (i.e.~on both the injection and extraction rates), and a \emph{fast} subphase, where it depends on just one of these control parameters.
All the analysis relies on a mean-field approximation, previously validated as a tool for studying dynamical transitions in the pure TASEP~\cite{PelizzolaPretti17}. 
The simplicity of the theory allows us to obtain many analytical results in the infinite-size limit, extending the approach developed in the balanced case~\cite{BPPZ18}.

Our main result is that the unbalanced Langmuir kinetics can change the nature of the dynamical transition and give rise to qualitatively new phenomena, which we have called (borrowing the term from equilibrium statistical physics) \emph{first-order-like} dynamical transitions. 
The latter term has different motivations. 
In the first place, we have observed that, depending on the model parameters, the singularity associated with the dynamical transition may also appear as a discontinuity in the first derivative of the relaxation rate with respect to either the injection or extraction rate. 
This is a novel feature, since both in the pure TASEP (which is exactly solvable)~\cite{deGierEssler05,deGierEssler06,deGierEssler08} and in some variants (including the TASEP with balanced Langmuir kinetics)~\cite{BottoPelizzolaPretti18,BPPZ18,PelizzolaPrettiPuccioni19} the relaxation rate only exhibits discontinuities in the second derivative (in such cases we speak of \emph{second-order-like} dynamical transitions). 
Furthermore, we have found that, when the relaxation rate exhibits first-order singularities, the whole spectrum of the mean-field relaxation matrix exhibits a non-trivial behaviour, being characterized (at finite size) by avoided crossings. 
Indeed, the first-order-like dynamical transition emerges as the infinite-size limit of one such avoided crossing. 
Still at finite size, we have characterized numerically the spatial structure of the slowest relaxation mode, i.e.~the eigenvector of the (mean-field) relaxation matrix corresponding to its smallest eigenvalue. 
Our results show that the first-order-like dynamical transition is accompanied by an abrupt structural change and in particular that the eigenvector exhibits a clearly bimodal shape, concurrently with the transition.
We argue that such a behavior may be loosely interpreted as a ``coexistence'' of two different relaxation modes, that is, a dynamical analogue of coexisting states at equilibrium first-order phase transitions. 

We have also obtained accurate numerical results about the finite-size scaling of the mean-field relaxation rate, showing that it approaches the asymptotic value with a power law. 
By comparison with the pure TASEP, we have conjectured that the exact value of the corresponding exponent is $2/3$ in the whole high- and low-density
phases. 

In conclusion, even though the present work is entirely based on a mean-field theory, the latter has proved to be reliable in the case of pure TASEP, where the dynamical transition is known exactly, and hence we expect that the first-order-like transition is not an artifact of the approximation. 
However, it may be worth trying to reproduce this phenomenon, as well as the scaling exponent, using complementary approaches. 
As shown in~\cite{ProemeBlytheEvans11}, Monte Carlo simulations are not a very efficient tool for studying dynamical transitions, so we plan to resort to a modified domain-wall theory and to extrapolation of finite-size results, as we have previously done for the Antal-Sch\"utz~\cite{BottoPelizzolaPretti18} and Katz-Lebowitz-Spohn~\cite{PelizzolaPrettiPuccioni19} models. 
Work is in progress along these lines.

\appendix

\section{Density and current profiles}
\label{sec:appendix_dens_prof}

In this appendix we report some details about the properties of the steady-state density and current profiles in the HD phase. 
Let us recall that, assuming fixed ${\Omega_\mathrm{A} > \Omega_\mathrm{D}}$, the HD-phase region is defined by the inequalities \eqref{eq:intervalli_valori_HD}. 
As discussed in~\cite{ParmeggianiFranoschFrey04}, in the continuum limit the density profile is the solution of a first-order differential equation, so that it can match at most one boundary condition (in the HD phase the right-boundary condition). 
As previously mentioned, this \emph{bulk solution} $\rho(h)$ can be expressed in terms of the Lambert $W$ function (in the HD phase the so-called \emph{main branch} $W_0$), in the following form
\begin{equation}
\frac{\rho(h) - \ell  \vphantom{\frac{1}{2}}}{\ell - \frac{1}{2}}
= 
{W}_{0} \left(
\frac{\oneminus{\beta} - \ell  \vphantom{\frac{1}{2}}}{\ell - \frac{1}{2}}
\, \exp \frac{\oneminus{\beta} - \ell - \Omegamedio \, \oneminus{h}}{\ell - \frac{1}{2}} 
\right)
\, ,
\label{eq:bulk_solution}
\end{equation}
where $\ell$ is the Langmuir density \eqref{eq:Langmuir_isotherm} and ${h \in [0,1]}$ is the position variable (${h = 0}$ and ${h = 1}$ denoting respectively the left and right boundaries).
Exploiting the fact that ${W_0\left( \xi \rme^{\xi} \right) = \xi}$ for all ${\xi \ge -1}$, and observing that \eqref{eq:intervallo_valori_beta} implies ${(\oneminus{\beta} - \ell)/(\ell - \frac{1}{2}) > -1}$, one can easily verify that 
\begin{equation}
\rho(1) = \oneminus{\beta}
\, ,
\label{eq:rhodiuno_uguale_betaprimo}
\end{equation}
i.e.~that the right-boundary condition is satisfied. 
Moreover, taking into account \eqref{eq:definizione_theta}, the left-boundary value of the bulk profile reads
\begin{equation}
\rho(0) = \vartheta
\, ,
\label{eq:rhodizero_uguale_theta}
\end{equation}
so that in general the left-boundary condition is not satisfied (the resulting mismatch is filled up by the \emph{boundary layer}). 

As previously observed for the balanced case~\cite{BPPZ18}, since our final goal is to investigate the eigenvalue problem for the relaxation matrix, it is convenient to perform the whole analysis at finite~$N$, in order to avoid dealing with infinite-dimensional operators. 
The basic difference with respect to our previous work~\cite{BPPZ18} is that here we do not have an analytical form for the bulk solution at finite $N$. 
Nevertheless, we can define the latter as a sequence ${(\dbulk_{n})}_{n=0}^{N+1}$ satisfying the steady-state equations \eqref{eq:p_steady_state}, that is 
\begin{equation}
\dbulk_{n} \left( \oneminus{\dbulk_{n+1}} + \omega_\mathrm{D} \right)
= \left( \dbulk_{n-1} + \omega_\mathrm{A} \right) \oneminus{\dbulk_{n}}
\qquad n = 1,\dots,N
\, ,
\label{eq:steady-state_bulk}
\end{equation}
with the boundary values matching precisely those of the continuum solution, i.e. 
\begin{subequations}
\label{eq:q_boundaries}
\begin{eqnarray}
\dbulk_{0}   & = \vartheta 
\label{eq:q_left}
\, , \\
\dbulk_{N+1} & = \oneminus{\beta} 
\label{eq:q_right}
\, .
\end{eqnarray}
\end{subequations}
Since the solution of the discrete boundary problem is unique~\cite{BPPZ18}, the above definition is unambiguous, and for large $N$ we can expect 
\begin{equation}
\dbulk_{n} \approx \rho \left( \textstyle \frac{n}{N+1} \right)
\qquad n = 0,\dots,N+1
\, ,
\label{eq:bulk_solution_discrete-continuous}
\end{equation}
where (from now on) the symbol $\approx$ means that the two sides of the equation can differ at most by an amount that vanishes for ${N \to \infty}$. 
Note that, at odds with \cite{BPPZ18}, in this paper we do not state rigorous bounds for the distance between the terms of approximate equalities, but it is understood that we have always verified numerically that such a distance actually vanishes for large $N$. 

After this step, we can proceed in analogy with the balanced case~\cite{BPPZ18}, by defining the \emph{detrended densities}
\begin{equation}
\ddetr_{n} \equiv \dfull_{n} - (\dbulk_{n} - \dbulk_{0})
\qquad n = 0,\dots,N+1  
\, ,
\label{eq:detrended_densities_definition}
\end{equation}
i.e.~the quantities obtained by subtracting from the local densities the non-uniform part of the bulk profile. 
From \eqref{eq:detrended_densities_definition} we have of course
\begin{subequations} 
\label{eq:pnboth_vs_rnboth}
\begin{eqnarray}
\dfull_{n} \hphantom{'}
= \ddetr_{n} \hphantom{'} + \dbulk_{n} \hphantom{'} - \dbulk_{0} \hphantom{'}
\, , \label{eq:pn_vs_rn} \\
\oneminus{\dfull_{n}} 
= \oneminus{\ddetr_{n}} + \oneminus{\dbulk_{n}} - \oneminus{\dbulk_{0}}
\, , \label{eq:pnprimo_vs_rnprimo} 
\end{eqnarray}
\end{subequations} 
which, plugged into the steady-state equations \eqref{eq:p_steady_state}, yield by simple algebra 
\begin{eqnarray}
\dbulk_{n} \left( \oneminus{\dbulk_{n+1}} + \omega_\mathrm{D} \right) 
+ \ddetr_{n} \oneminus{\ddetr_{n+1}} 
- \left( \dbulk_{n} - \dbulk_{0} \right) \ddetr_{n+1}
- \left( \ddetr_{n} - \dbulk_{0} \right) \left( \dbulk_{n+1} - \omega_\mathrm{D} \right) 
= \nonumber \\ 
\left( \dbulk_{n-1} + \omega_\mathrm{A} \right) \oneminus{\dbulk_{n}} 
+ \ddetr_{n-1} \oneminus{\ddetr_n} 
- \left( \dbulk_{n} - \dbulk_{0} \right) \ddetr_{n-1}
- \left( \ddetr_{n} - \dbulk_{0} \right) \left( \dbulk_{n-1} + \omega_\mathrm{A} \right) 
\end{eqnarray}
for all $n=1,\dots,N$. 
We notice that the first terms on both sides cancel out because of \eqref{eq:steady-state_bulk}, so that we get 
\begin{eqnarray}
\ddetr_{n} \oneminus{\ddetr_{n+1}} 
- \ddetr_{n-1} \oneminus{\ddetr_n} 
= & \left( \dbulk_{n} - \dbulk_{0} \right) \left( \ddetr_{n+1} - \ddetr_{n-1} \right) 
+ \nonumber \\
& \left( \ddetr_{n} - \dbulk_{0} \right) 
\left( \dbulk_{n+1} - \dbulk_{n-1} - \omega_\mathrm{A} - \omega_\mathrm{D} \right) 
\label{eq:SteadyStateMF_detrended}
\end{eqnarray}
still for all $n=1,\dots,N$. 
Because of \eqref{eq:detrended_densities_definition}, \eqref{eq:q_boundaries} and \eqref{eq:p_boundaries}, the boundary conditions become 
\begin{subequations} 
\begin{eqnarray}
\label{eq:SteadyStateMF_detrended_left}
\ddetr_{0} & = {p}_{0} & = \alpha
\, , \\
\label{eq:SteadyStateMF_detrended_right}
\ddetr_{N+1} & = \dbulk_{0} & = \vartheta
\, . 
\end{eqnarray}
\end{subequations} 
Now, since we know from numerics that in the HD phase the local density $\dfull_{n}$ approaches exponentially the bulk solution, with the characteristic length of the exponential remaining finite as $N$ grows to infinity, we can expect that the detrended density profile will be similar to the density profile of an ``effective pure TASEP''. 
One can easily verify that the consequences of this conjecture are fully consistent. 
In particular, assuming that $\ddetr_{n}$ tends to $\dbulk_{0}$ exponentially upon increasing $n$, the difference ${\ddetr_{n+1} - \ddetr_{n-1}}$, appearing in the right-hand side of \eqref{eq:SteadyStateMF_detrended}, would be significantly different from zero only up to finite ${n}$. 
Yet, in this region the factor ${\dbulk_{n} - \dbulk_{0}}$ turns out be of order $1/N$, because, according to \eqref{eq:bulk_solution_discrete-continuous}, $\dbulk_{n}$ can be regarded as a discretization of the continuous function $\rho(h)$.
Similar arguments apply to the last term in \eqref{eq:SteadyStateMF_detrended}, leading to the conclusion that the whole right-hand side of \eqref{eq:SteadyStateMF_detrended} can be neglected, and the detrended densities satisfy 
\begin{equation}
\ddetr_{n} \oneminus{\ddetr_{n+1}} \approx \mathrm{constant}
\qquad n = 0,1,\dots,N
\, .
\label{eq:SteadyStateMF_detrended_approx}
\end{equation}
The latter are indeed pure-TASEP mean-field equations (check \eqref{eq:p_steady_state} with ${\omega_\mathrm{A} = \omega_\mathrm{D} = 0}$), with a renormalized right-boundary condition, namely \eqref{eq:SteadyStateMF_detrended_right}. 
The resulting physical picture is as well analogous to that of the balanced case. 
The local density $\dfull_{n}$ undergoes variations of order $1$ just over a finite number of nodes. 
In such a small region (boundary layer), the effect of Langmuir kinetics becomes negligible for large $N$ and, as a consequence, the system behaves there as a pure TASEP with a bulk density adjusted to match the local bulk density of the TASEP-LK in that region. 
In particular, in the HD phase the boundary layer is on the left, so that the effective pure TASEP is one with a right-boundary condition renormalized to the \emph{left}-boundary value of the bulk solution, that is ${\dbulk_{0} = \vartheta}$.
Let us observe that equations \eqref{eq:SteadyStateMF_detrended_approx} are more and more accurate as ${N \to \infty}$, so that it is natural to define an infinite sequence ${(\ddlim_{n})}_{n=0}^{\infty}$, being a solution of \eqref{eq:SteadyStateMF_detrended_approx} taken as an equality, namely 
\begin{equation}
\ddlim_{n} \oneminus{\ddlim_{n+1}}
= \vartheta \oneminus{\vartheta}
\qquad n=0,1,2,\dots
\, ,
\label{eq:proposizione_eta}
\end{equation}
where the constant (right-hand side) reflects the fact that $\ddlim_{n}$ should approach the bulk value $\vartheta$. 
Choosing also the left-boundary condition according to \eqref{eq:SteadyStateMF_detrended_left}, that is 
\begin{equation}
\ddlim_{0} = \alpha
\, ,
\label{eq:proposizione_eta_0}
\end{equation}
in the limit of large $N$ we expect that 
\begin{equation} 
\ddetr_{n} \approx \ddlim_{n}
\qquad n=0,\dots,N+1
\, .
\label{eq:rn_approx_sn}
\end{equation}
Let us recall that the closed-form solution of \eqref{eq:proposizione_eta} with the ``initial'' condition \eqref{eq:proposizione_eta_0} can be written as~\cite{BPPZ18}
\begin{equation}
\ddlim_{n} = \vartheta + \psi_{n} 
\qquad n = 0,1,2,\dots
\, ,
\label{eq:definizione_s}
\end{equation} 
where
\begin{equation}
\psi_{n} \equiv 
(\vartheta - \oneminus{\vartheta}) 
\left\{ \left[ 1 - 
\frac{\alpha - \vartheta\hphantom{'}}{\alpha - \oneminus{\vartheta}}
\, \left( \frac{\oneminus{\vartheta}}{\vartheta} \right)^{\!\! n \,} \right]^{-1} - 1 \right\}
\, ,
\label{eq:definizione_psi}
\end{equation} 
which is in fact equivalent to the one reported in \cite{DerridaDomanyMukamel92}.
Let us also note that \eqref{eq:definizione_theta} along with the parameter bound \eqref{eq:intervallo_valori_beta} entail
\begin{equation}
0 < \frac{\oneminus{\vartheta}}{\vartheta} < 1
\, ,
\label{eq:intervallo_valori_gamma}
\end{equation}
which in turn shows that ${\psi_{n} \to 0}$ (exponentially) for ${n \to \infty}$. 
In the end, taking into account \eqref{eq:pn_vs_rn}, \eqref{eq:q_left}, \eqref{eq:bulk_solution_discrete-continuous}, \eqref{eq:rn_approx_sn} and \eqref{eq:definizione_s}, we can state that the stationary density profile of the HD phase is
\begin{equation}
{p}_{n} =  \dbulk_{n} + \ddetr_{n} - \dbulk_{0}
\approx \rho \left ( \textstyle \frac{n}{N+1} \right) + \psi_{n}
\qquad n = 0,\dots,N+1
\, ,
\end{equation}
where $\rho(h)$ and $\psi_{n}$ are defined respectively by \eqref{eq:bulk_solution} and \eqref{eq:definizione_psi}, together with \eqref{eq:definizione_theta}. 

It is possible to show that the above arguments entail relevant consequences regarding also the current profile, in particular that the latter is very close (for large ${N}$) to the current profile corresponding to the bulk solution alone (which we shall denote as \emph{bulk current}). 
In fact, using \eqref{eq:pnboth_vs_rnboth}, we can write by simple algebra 
\begin{eqnarray}
\dfull_{n} \oneminus{\dfull_{n+1}} - \dbulk_{n} \oneminus{\dbulk_{n+1}} 
= \left( \ddetr_{n} \oneminus{\ddetr_{n+1}} - \dbulk_{0} \oneminus{\dbulk_{0}} \right) 
& - \left( \dbulk_{n} - \dbulk_{0} \right) \left( \ddetr_{n+1} - \dbulk_{0} \right)
\nonumber \\
& - \left( \dbulk_{n+1} - \dbulk_{0} \right) \left( \ddetr_{n} - \dbulk_{0} \right)
\end{eqnarray}
for all ${n=0,\dots,N}$.
The first term on the right-hand side is negligible because of \eqref{eq:rn_approx_sn} and \eqref{eq:proposizione_eta}, along with \eqref{eq:q_left}. 
The subsequent terms can be neglected as well, because of the same arguments used for \eqref{eq:SteadyStateMF_detrended}, so that in the end we obtain the expected result
\begin{equation}
{J}_{n} = \dfull_{n} \oneminus{\dfull_{n+1}}
\approx \dbulk_{n} \oneminus{\dbulk_{n+1}}
\approx \rho \left ( \textstyle \frac{n}{N+1} \right) 
\oneminus{\rho} \left ( \textstyle \frac{n}{N+1} \right) 
\qquad n = 0,\dots,N
\, ,
\label{eq:current_discrete-continuous}
\end{equation}
where the latter approximate equality follows from \eqref{eq:bulk_solution_discrete-continuous} together with the continuity of $\rho(h)$. 
One can easily argue that in the HD phase, where the bulk density is always larger than $1/2$, the maximum current corresponds to the minimum bulk density, as $\rho\oneminus{\rho}$ is a decreasing function of $\rho$ for ${\rho > 1/2}$.
As a consequence, keeping in mind figure~\ref{fig:density_profiles}, the maximum current may occur either at the left or at the right boundary (respectively taking values $\vartheta\oneminus{\vartheta}$ or $\beta\oneminus{\beta}$) depending on whether ${\beta < \oneminus{\ell}}$ or ${\beta > \oneminus{\ell}}$. 
This last result will be invoked several times in the forthcoming section.

\section{Relaxation rates}
\label{sec:appendix_eigenvalues}

In section~\ref{sec:model} we have seen that the relaxation rates are the eigenvalues of a tridiagonal symmetric matrix, which we now call $A$, mapping the generic vector ${u \equiv (u_1,\dots,u_N)}$ to ${Au}$. 
The expression in components of the latter vector is 
\begin{equation}
{(Au)}_{n} \equiv {a}_{n} {u}_{n} 
- \sqrt{{J}_{n}  } \, {u}_{n+1}  
- \sqrt{{J}_{n-1}} \, {u}_{n-1}  
\qquad n=1,\dots,N
\, ,
\label{eq:definizione_A}
\end{equation}
where the diagonal term ${a}_{n}$ is defined by \eqref{eq:definizione_a}, ${{J}_{n} = \dfull_{n} \oneminus{\dfull_{n+1}}}$ is the local steady-state current and ${{u}_{0} \equiv {u}_{N+1} \equiv 0}$. 
The slowest relaxation rate is the smallest eigenvalue of $A$, which we denote (emphasizing the dependence on $N$) by ${\lambda_{\min}}^{(N)}$. 
Our goal in this appendix is the analytical derivation of the asymptotic value ${\lambda_{\min}}^{(\infty)}$, still in the HD phase. 
The basic idea is to find suitable upper- and lower-bounds, which may coincide in the limit ${N \to \infty}$. 

\subsection{Upper-bounds}

According to \emph{Courant's minimax principle}, we can state the inequality 
\begin{equation}
{\lambda_{\min}}^{(N)} \leq (u,Au)
\, 
\label{eq:Courant-type_bound}
\end{equation}
for any vector ${u \in \mathbb{R}^{N}}$ such that ${\|{u}\| = 1}$, where ${(u,v) \equiv \sum_{n=1}^{N} {u}_{n} {v}_{n}}$ is the usual Euclidean scalar product and ${\|{u}\| \equiv \sqrt{(u,u)}}$ is the corresponding norm. 
From now on we shall always assume that $u$ is a vector with the above property (i.e.~a \emph{normalized} vector), or equivalently that  $u_1,\dots,u_N$ are real numbers such that ${\sum_{n=1}^{N} {{u}_{n}}^{2} = 1}$. 
Taking into account \eqref{eq:definizione_A}, we have in general 
\begin{equation}
(u,Au) = \sum_{n=1}^{N} {a}_{n} {{u}_{n}}^2
- 2 \sum_{n=1}^{N-1} \sqrt{{J}_{n}} \, {u}_{n} {u}_{n+1}
\, ,
\label{eq:uau_1}
\end{equation}
where, plugging \eqref{eq:pn_vs_rn} into \eqref{eq:definizione_a}, 
\begin{equation}
{a}_{n} 
= 1 - \left( \ddetr_{n+1} - \ddetr_{n-1} \right) 
- \left( \dbulk_{n+1} - \dbulk_{n-1} - \omega_\mathrm{A} - \omega_\mathrm{D} \right) 
\, .
\label{eq:a_n}
\end{equation}
Since we are interested in the limit ${N \to \infty}$, the quadratic form $(u,Au)$ can be manipulated, neglecting any term that vanishes for large $N$. 
In particular, as previously argued in \ref{sec:appendix_dens_prof}, we can neglect the last term in the right-hand side of \eqref{eq:a_n}, and replace $\ddetr_{n}$ with $\ddlim_{n}$, so that  
\begin{equation}
{a}_{n} 
\approx 1 - \left( \ddlim_{n+1} - \ddlim_{n-1} \right) 
\label{eq:a_n_approx}
\end{equation}
and therefore
\begin{equation}
(u,Au) \approx 1 
- \sum_{n=1}^{N} \left( \ddlim_{n+1} - \ddlim_{n-1} \right) {{u}_{n}}^2
- 2 \sum_{n=1}^{N-1} \sqrt{{J}_{n}} \, {u}_{n} {u}_{n+1}
\, .
\label{eq:uau_2}
\end{equation}

\paragraph{Upper-bound (i)}  

For a given integer ${M > 1}$, let us define the sequence
\begin{equation}
{y}_{n} \equiv 
\sqrt{\frac{2}{M}} \, \sin \frac{\pi n}{M} 
\, ,
\label{eq:definizione_w}
\end{equation}
for which the following properties can be easily verified
\begin{subequations} 
\begin{eqnarray}
\sum_{n=1}^{M-1} {{y}_{n}}^{2} = 1
\, , \label{eq:somma_w2} \\
\sum_{n=1}^{M-2} {y}_{n} {y}_{n+1} = \cos \frac{\pi}{M}
\, . \label{eq:somma_ww}
\end{eqnarray}
\end{subequations} 
In particular, for any ${M \in \{ 2,\dots,N+1 \}}$, \eqref{eq:somma_w2} allows one to choose a normalized trial-vector $u$ as 
\begin{equation}
{u}_{n} \equiv \cases{
	{y}_{n} & if ${n \le M}$ \\
	0       & if ${n \ge M}$
}
\qquad n=1,\dots,N
\, .
\label{eq:definizione_u_1}
\end{equation}
We take care of choosing $M$ in such a way that, for ${N \to \infty}$, one has ${M \to \infty}$ yet ${M/N \to 0}$. 
This means that, for large $N$, the vector $u$ will have at the same time a large number and a vanishing fraction of nonzero components, being concentrated at low $n$ values (that is at the left boundary of the system). 
Let us now consider the right-hand side of \eqref{eq:uau_2}. 
From \ref{sec:appendix_dens_prof} we know that the difference ${\ddlim_{n+1} - \ddlim_{n-1}}$ vanishes exponentially upon increasing $n$, so that for large $N$ (and thence large $M$) we can write 
\begin{equation}
\sum_{n=1}^{N} \left( \ddlim_{n+1} - \ddlim_{n-1} \right) {{u}_{n}}^2
\approx 0
\, .
\label{eq:prima_somma_circa_zero}
\end{equation}
Regarding the other sum, we can argue 
\begin{equation}
\sum_{n=1}^{N-1} \sqrt{{J}_{n}} \, {u}_{n} {u}_{n+1}
= \sum_{n=1}^{M-2} \sqrt{{J}_{n}} \, {y}_{n} {y}_{n+1}
\approx \sqrt{\vphantom{{J}_{n}} \vartheta \oneminus{\vartheta}} 
\, \sum_{n=1}^{M-2} {y}_{n} {y}_{n+1}
\, ,
\end{equation}
where the exact equality descends immediately from \eqref{eq:definizione_u_1} and   the approximate one can be explained as follows. 
According to \eqref{eq:current_discrete-continuous}, for large $N$ the local current ${J}_{n}$ can be treated as a continuous function of the position variable ${z \equiv n/(N+1)}$. 
Since ${M/N \to 0}$, we have ${z \to 0}$ for all $n$ values included in the sum, then ${J}_{n}$ can be replaced by the left-boundary value ${\rho(0)\oneminus{\rho}(0) = \vartheta \oneminus{\vartheta}}$, where this last equality obviously descends from \eqref{eq:rhodizero_uguale_theta}. 
Taking into account \eqref{eq:somma_ww} along with ${M \to \infty}$, we then get 
\begin{equation}
(u,Au) \approx 1 - 2 \sqrt{\vphantom{\beta} \vartheta \oneminus{\vartheta}} 
\, ,
\label{eq:uau_3}
\end{equation}
whereas \eqref{eq:Courant-type_bound} finally gives 
\begin{equation}
{\lambda_{\min}}^{(\infty)} \leq 1 - 2 \sqrt{\vphantom{\beta} \vartheta \oneminus{\vartheta}} 
\, .
\label{eq:upperbound_1}
\end{equation}

\paragraph{Upper-bound (ii)}  

This bound is very similar to the previous one, in that we choose a trial-vector $u$ being concentrated at the right (rather than left) boundary, namely
\begin{equation}
{u}_{n} \equiv \cases{
	0           & if ${n \le N+1-M}$ \\
	{y}_{N+1-n} & if ${n \ge N+1-M}$
}
\qquad n=1,\dots,N
\, ,
\label{eq:definizione_u_2}
\end{equation}
with ${y}_{n}$ still defined by \eqref{eq:definizione_w}.
Considering the right-hand side of \eqref{eq:uau_2}, we can first observe that \eqref{eq:prima_somma_circa_zero} still holds. 
Regarding the other sum, we can argue 
\begin{equation}
\sum_{n=1}^{N-1} \sqrt{{J}_{n}} \, {u}_{n} {u}_{n+1}
= \sum_{n=1}^{M-2} \sqrt{{J}_{N+1-n}} \, {y}_{n} {y}_{n+1}
\approx \sqrt{\beta \oneminus{\beta}} \, \sum_{n=1}^{M-2} {y}_{n} {y}_{n+1}
\, .
\end{equation}
In analogy with upper-bound~(i), the exact equality follows from \eqref{eq:definizione_u_2}, whereas the approximate one descends from the fact that, for all $n$ values included in the sum, the local current ${J}_{N+1-n}$ is well approximated (in the usual sense) by the right-boundary value ${\rho(1)\oneminus{\rho}(1) = \beta \oneminus{\beta}}$, the last equality obviously descending from \eqref{eq:rhodiuno_uguale_betaprimo}.  
Still taking into account \eqref{eq:somma_ww} with ${M \to \infty}$, we get 
\begin{equation}
(u,Au) \approx 1 - 2 \sqrt{\beta \oneminus{\beta}}
\, ,
\label{eq:uau_4}
\end{equation}
and by \eqref{eq:Courant-type_bound} we finally obtain 
\begin{equation}
{\lambda_{\min}}^{(\infty)} \leq 1 - 2 \sqrt{\beta \oneminus{\beta}} 
\, .
\label{eq:upperbound_2}
\end{equation}

\paragraph{Upper-bound (iii)} 

The above two bounds do not depend on $\alpha$ and hold in principle with no restrictions on $\alpha$, except ${\alpha > \oneminus{\vartheta}}$, required in order to stay within the HD-phase region (as it is always understood). 
Looking at the numerical results (subsection~\ref{subsec:spectrum}), one obviously expects that in the low-$\alpha$ range (i.e.~below the dynamical transition) these cannot be good bounds, because in that regime the smallest eigenvalue gets much lower than its plateau value and strongly dependent on $\alpha$.
The basic idea, previously exploited in~\cite{BPPZ18}, is that, in order to obtain good bounds, we have to choose ${u}$ in \eqref{eq:Courant-type_bound} as close as possible to the actual eigenvector. 
Such an argument underlies definition \eqref{eq:definizione_v} for the family of sequences ${v}_{n}(x)$ (parameterized by the real variable $x$). 
Based on the above, we define 
\begin{equation}
{u}_{n} \equiv \frac{{v}_{n}({x}_{*})}{\sqrt{\sum_{k=1}^{N} {{v}_{k}({x}_{*})}^{2}}}
\qquad n=0,1,2,\dots
\, ,
\label{eq:definizione_u}
\end{equation}
where $x_*$ is a function of the model parameters, defined as in subsection~\ref{subsec:analytics}.\footnote{As proved in \cite{BPPZ18}, ${x}_{*}$ represents the smallest real value such that the sequence ${v}_{n}(x_*)$ is strictly positive, except ${v}_{0}(x_*)$ being zero by definition.} 
Note that \eqref{eq:definizione_u} obviously entails ${\sum_{n=1}^{N} {{u}_{n}}^{2} = 1}$ for any $N$, as required in order to use ${u_1,\dots,u_N}$ for Courant-type bounds in the form \eqref{eq:Courant-type_bound}, whereas the reason for the specific choice of $x_*$ will be clear in the following. 
Taking into account \eqref{eq:definizione_v} we have 
\begin{equation}
\frac{ \ddlim_{n+1} - \ddlim_{n-1} }{ \sqrt{\vartheta \oneminus{\vartheta}} } \, {u}_{n} 
= 2 {x}_{*} {u}_{n} - {u}_{n+1} - {u}_{n-1}
\qquad n=1,2,\dots
\, .
\label{eq:successione_u}
\end{equation}
Let us multiply both sides of the above equation by ${{u}_{n}}$ and sum over ${n=1,\dots,N}$. 
Keeping in mind that \eqref{eq:definizione_u} also implies ${{u}_{0} = 0}$, we obtain
\begin{equation}
\sum_{n=1}^{N} 
\frac{ \ddlim_{n+1} - \ddlim_{n-1} }{ \sqrt{\vartheta \oneminus{\vartheta}} } 
\, {{u}_{n}}^2
= 2 {x}_{*} - 2 \sum_{n=1}^{N-1} {u}_{n} {u}_{n+1} - {u}_{N} {u}_{N+1}
\, .
\label{eq:uau_passaggio_intermedio}
\end{equation}
At this point we need to recall that, as proven in \cite{BPPZ18}, if ${x_*>1}$ then the sequence $v_n(x_*)$ can be upper-bounded (up to a positive constant factor) by ${\zeta(x_*)}^{n}$, where ${\zeta(x_*) < 1}$ follows easily from \eqref{eq:definizione_z}. 
Also taking into account that ${\sum_{k=1}^{N} {{v}_{k}({x}_{*})}^{2} \ge {{v}_{1}({x}_{*})}^{2} = 1}$, we see that ${u}_{n}$ admits the same upper-bound, which immediately entails that the last term in the right-hand side of \eqref{eq:uau_passaggio_intermedio} can be dropped. 
Then, plugging the remainder of \eqref{eq:uau_passaggio_intermedio} into \eqref{eq:uau_2}, we obtain
\begin{equation}
(u,Au) \approx 1 
- 2 {x}_{*} \sqrt{\vphantom{J_n} \vartheta \oneminus{\vartheta}} 
- 2 \sum_{n=1}^{N-1} 
\left( \! \sqrt{{J}_{n}} - \sqrt{\vphantom{J_n} \vartheta \oneminus{\vartheta}} \, \right) 
{u}_{n} {u}_{n+1}
\, .
\label{eq:uau_5}
\end{equation}
In the right-hand side of this last equation we can see that, due to the usual argument, the difference in brackets is of order $n/N$. 
As a consequence, still taking into account that $u_n$ is bounded by ${{\zeta(x_*)}^{n}}$, for ${N \to \infty}$ the whole sum turns out to be of order $1/N$, and can be neglected. 
We thus obtain 
\begin{equation}
(u,Au) \approx 1 - 2 {x}_{*} 
\sqrt{\vphantom{\beta} \vartheta \oneminus{\vartheta}} 
\, 
\label{eq:uau_6}
\end{equation}
and thence
\begin{equation}
{\lambda_{\min}}^{(\infty)} \leq 1 - 2 {x}_{*} 
\sqrt{\vphantom{\beta} \vartheta \oneminus{\vartheta}} 
\, .
\label{eq:upperbound_3}
\end{equation}
Let us stress the fact that this last bound holds only assuming that ${x_* > 1}$, which corresponds by definition to ${\alpha < \alpha_\mathrm{c}}$, as discussed in subsection~\ref{subsec:analytics}.

\paragraph{Overall upper-bound}

In conclusion, let us consider the above three bounds together. 
Let us first observe that upper-bound~(i) can be viewed as the analogue of (iii) with ${x_* = 1}$. 
Consequently, we can state that \eqref{eq:upperbound_3} holds in fact for ${x_* \ge 1}$, that is for any admissible value of $x_*$, that is with no restrictions on $\alpha$. 
Furthermore, since all bounds hold simultaneously, we can put together the last statement with upper-bound~(ii), which leads to the overall bound 
\begin{equation}
{\lambda_{\min}}^{(\infty)} 
\le 1 - 2 \max \left\{ 
{x}_{*} \sqrt{ \vphantom{\beta} \vartheta \oneminus{\vartheta} } , 
\sqrt{ \beta \oneminus{\beta} } 
\, \right\}
\, .
\label{eq:upperbound_all}
\end{equation}

\subsection{Lower-bounds}

As in our previous work~\cite{BPPZ18}, we deal with lower-bounds by means of an argument closely related to \emph{Gershgorin's circle theorem}. 
Let us consider an infinite sequence ${w_0,w_1,w_2,\dots}$ of positive real numbers (except $w_0$, possibly being zero) and let ${u \in \mathbb{R}^{N}}$ be an eigenvector of $A$ corresponding to the smallest eigenvalue ${\lambda_{\min}}^{(N)}$. 
The vector $u$ can be chosen in such a way that ${u_{m} = w_{m}}$ for some ${m \in \{1,\dots,N\}}$ and ${|u_{n}| \leq w_{n}}$ for all other ${n \neq m}$ (including ${n = 0}$ and ${n = N+1}$, since by definition ${u_{0} \equiv u_{N+1} \equiv 0}$). 
In practice, given $u$ with an arbitrary normalization, one can take $m$ to be an index where the maximum of ${\{|u_{1}|/w_1,\dots,|u_{N}|/w_{N}\}}$ is attained, and then renormalize $u$ dividing by ${u_{m}/w_{m}}$. 
From this argument, we have in particular that ${u_{m} \neq 0}$, so that we can write \begin{equation}
{\lambda_{\min}}^{(N)} 
=
\frac{{(Au)}_{m}}{{u}_{m}}
= 
{a}_{m} 
- \sqrt{{J}_{m}  } \, \frac{{u}_{m+1}}{{u}_{m}} 
- \sqrt{{J}_{m-1}} \, \frac{{u}_{m-1}}{{u}_{m}} 
\, . 
\end{equation}
Then, using ${u_{m} = w_{m} > 0}$ and ${u_{m \pm 1} \leq w_{m \pm 1}}$, we easily get the bound
\begin{equation}
{\lambda_{\min}}^{(N)} 
\ge
{a}_{m} 
- \sqrt{{J}_{m}  } \, \frac{{w}_{m+1}}{{w}_{m}} 
- \sqrt{{J}_{m-1}} \, \frac{{w}_{m-1}}{{w}_{m}} 
\, . 
\end{equation}
At this point, in general one cannot foresee the proper index ${m}$, so that we are forced to choose the worst case, that is 
\begin{equation}
{\lambda_{\min}}^{(N)} 
\geq 
\min_{n=1\vphantom{|}}^{N} \left\{ 
{a}_{n} 
- \sqrt{{J}_{n}  } \, \frac{{w}_{n+1}}{{w}_{n}} 
- \sqrt{{J}_{n-1}} \, \frac{{w}_{n-1}}{{w}_{n}} 
\right\}
\, .
\label{eq:Gershgorin-type_bound}
\end{equation}
As for the upper-bounds, we shall only be interested in the limit ${N \to \infty}$, and therefore we can treat the right-hand side of \eqref{eq:Gershgorin-type_bound}, neglecting contributions that go to zero in such a limit. 
In particular we can express $a_{n}$ by \eqref{eq:a_n_approx} and, because of the usual argument that $J_{n}$ behaves like a continuous function of ${n/N}$, we can safely write ${J_{n-1} \approx J_{n}}$. 
We thus obtain 
\begin{equation}
{\lambda_{\min}}^{(N)} 
\gtrapprox
1 - \max_{n=1\vphantom{|}}^{N\vphantom{|}} \Delta_{n} 
\, ,
\label{eq:Gershgorin-type_bound_approx}
\end{equation}
where we have defined
\begin{equation}
\Delta_{n}
\equiv 
\ddlim_{n+1} - \ddlim_{n-1}
+ \sqrt{{J}_{n}} \, \frac{{w}_{n+1} + {w}_{n-1}}{{w}_{n}} 
\qquad n=1,\dots,N 
\, 
\label{eq:definizione_Delta}
\end{equation}
and where the symbol $\gtrapprox$ means that either the left-hand side is larger than the right-hand side or it may be smaller, but at most by an amount which vanishes for ${N \to \infty}$. 

The difficulties one encounters in applying this sort of bound turn out to be considerably different, depending on the parameter values. 
In particular, as far as the HD phase is concerned, we can distinguish three different cases of increasing complexity.

\paragraph{Case ${\alpha \ge \vartheta}$.}

The parameter region ${\alpha \geq \vartheta}$ turns out to be quite simple, because, according to \eqref{eq:definizione_s} and \eqref{eq:definizione_psi}, in this case the sequence $\ddlim_{n}$ is non-increasing, which entails ${\ddlim_{n+1} - \ddlim_{n-1} \leq 0}$. 
As a consequence, \eqref{eq:definizione_Delta} with the simple choice ${w}_{n} \equiv \mathrm{constant}$ (which turns out to coincide with the usual Gershgorin bound) immediately gives 
\begin{equation}
\Delta_{n}
\le 
2 \sqrt{{J}_{n}} 
\qquad n=1,\dots,N 
\, .
\label{eq:Gershgorin-type_bound_alfagrandi}
\end{equation}
Then, by \eqref{eq:Gershgorin-type_bound_approx} and \eqref{eq:Gershgorin-type_bound_alfagrandi}, taking the limit ${N \to \infty}$ and exploiting the usual fact that ${J}_{n}$ can be approximated by \eqref{eq:current_discrete-continuous} (and therefore that the maximum current value may be either ${\vartheta \oneminus{\vartheta}}$ or ${\beta \oneminus{\beta}}$), we obtain
\begin{equation}
{\lambda_{\min}}^{(\infty)} 
\ge 1 - 2 \max \left\{ 
\! \sqrt{ \vphantom{\beta} \vartheta \oneminus{\vartheta} } , 
\sqrt{ \beta \oneminus{\beta} } 
\, \right\}
\, .
\label{eq:lowerbound_largealpha}
\end{equation}
Comparing this lower-bound with the upper-bound \eqref{eq:upperbound_all}, and recalling that for ${\alpha \ge \vartheta}$ we always have ${x_* = 1}$, we easily conclude that in this region the bound is tight, i.e.~it holds as an equality.

\paragraph{Case ${\alpha_\mathrm{c} \le \alpha < \vartheta}$.}

Following \cite{BPPZ18}, let us choose 
\begin{equation}
{w}_{n} \equiv {v}_{n}(x_*)
\qquad n=0,1,2,\dots 
\, , 
\label{eq:definizione_w_per_lowerbound_semplice}
\end{equation}
where $v_n(x)$ still denotes the family of sequences defined by \eqref{eq:definizione_v}. 
Note that this choice is feasible because in \cite{BPPZ18} we proved that $v_n(x_*)$ is strictly positive (except ${v_0(x_*) = 0}$). 
With the above definition, \eqref{eq:definizione_Delta} obviously reads
\begin{equation}
\Delta_{n} 
= 
\ddlim_{n+1} - \ddlim_{n-1}
+ \sqrt{{J}_{n}} \, \frac{{v}_{n+1} + {v}_{n-1}}{{v}_{n}} 
\qquad n=1,\dots,N 
\, ,
\label{eq:Delta_euno}
\end{equation}
where $v_n$ is used as a shorthand for $v_n(x_*)$, as it will be from now on. 
From \eqref{eq:definizione_v} we get 
\begin{equation}
\ddlim_{n+1} - \ddlim_{n-1}
= 
\sqrt{ \vphantom{\beta} \vartheta \oneminus{\vartheta} }
\left[
2 {x}_{*} 
- \frac{{v}_{n+1} + {v}_{n-1}}{{v}_{n}} 
\right]
\qquad n=1,2,\dots 
\, ,
\label{eq:differenza_s-s}
\end{equation}
which, plugged into \eqref{eq:Delta_euno}, gives
\begin{equation}
\Delta_{n} 
= 
2 {x}_{*} \sqrt{ \vphantom{\beta} \vartheta \oneminus{\vartheta} }
+ \left( \! 
\sqrt{{J}_{n}} 
- \sqrt{ \vphantom{\beta} \vartheta \oneminus{\vartheta} } \, 
\right) 
\frac{{v}_{n+1} + {v}_{n-1}}{{v}_{n}} 
\qquad n=1,\dots,N 
\, .
\label{eq:Delta_edue}
\end{equation}
Now, according to \eqref{eq:definizione_s} and \eqref{eq:definizione_psi}, the hypothesis ${\alpha < \vartheta}$ implies ${\ddlim_{n+1} - \ddlim_{n-1} > 0}$, so that from \eqref{eq:differenza_s-s} we also get 
\begin{equation}
\frac{{v}_{n+1} + {v}_{n-1}}{{v}_{n}} 
<
2 {x}_{*} 
\qquad n=1,2,\dots 
\, .
\label{eq:disuguaglianza_rapp-v}
\end{equation}
Plugging this last inequality into \eqref{eq:Delta_edue}, one can easily deduce
\begin{equation}
\Delta_{n} 
\le 
2 {x}_{*} \max \left\{ \!
\sqrt{ \vphantom{\beta} \vartheta \oneminus{\vartheta} } , 
\sqrt{ \vphantom{\beta} {J}_{n} }
\, \right\}
\qquad n=1,\dots,N 
\, .
\end{equation}
Hence \eqref{eq:Gershgorin-type_bound_approx} in the limit ${N \to \infty}$ and the usual argument for $J_n$ (i.e.~that its maximum value may be either ${\vartheta \oneminus{\vartheta}}$ or ${\beta \oneminus{\beta}}$) finally yield 
\begin{equation}
{\lambda_{\min}}^{(\infty)} 
\ge 
1 - 2 {x}_{*} \max \left\{ \!
\sqrt{ \vphantom{\beta} \vartheta \oneminus{\vartheta} } , 
\sqrt{ \beta \oneminus{\beta} }
\, \right\}
\, .
\end{equation}
Comparing the above lower-bound with \eqref{eq:upperbound_all}, we can see that even in this region both bounds are tight, since for ${\alpha \ge \alpha_\mathrm{c}}$ we still have ${x_* = 1}$.

More precisely, let us observe that, limited to the low-$\beta$ region ${\beta \le \oneminus{\ell}}$, these bounds turn out to be actually tight for all ${\alpha < \vartheta}$ (within the HD-phase border ${\alpha > \oneminus{\vartheta}}$). 
Indeed, for ${\beta \le \oneminus{\ell}}$ we know that the current profile is non-increasing (as in the balanced case~\cite{BPPZ18}), so that in particular ${\vartheta \oneminus{\vartheta} \ge \beta \oneminus{\beta}}$. 
As a consequence, the term ${\beta \oneminus{\beta}}$ (right-boundary current) can be dropped, and the two bounds lead to equality \eqref{eq:slowest_rate_second}, regardless of the $\alpha$ value.

\paragraph{Case ${\alpha < \alpha_\mathrm{c}}$.}

We still recall that, as discussed in subsection~\ref{subsec:analytics}, ${\alpha < \alpha_\mathrm{c}}$ corresponds by definition to ${x_* > 1}$. 
In this hypothesis, building on the results of \cite{BPPZ18}, it is possible to prove that
\begin{equation}
\lim_{n \to \infty} \frac{{v}_{n+1}(x_*)}{{v}_{n}(x_*)}
= \zeta(x_*) < 1
\, ,
\label{eq:limite_rapporto_v}
\end{equation}
where ${\zeta(x_*) < 1}$ follows immediately from \eqref{eq:definizione_z}. 
A proof of the above statement, which turns out to be of use in the following discussion, will be given afterwards. 
For the sequence $w_n$, let us consider the following expression  
\begin{equation}
{w}_{n} \equiv \cases{
	{v}_{n}(x_*)                                              & if ${n \le L}$ \\
	{v}_{L}(x_*)                                              & if ${n \ge M}$ \\
	\phi_{n} {v}_{L}(x_*) + \oneminus{\phi_{n}} {v}_{n}(x_*)  & if ${L \le n \le M}$ 
}
\ \ n=0,1,2,\dots
\, ,
\label{eq:definizione_w_per_lowerbound}
\end{equation}
where $L$ and $M$ are positive integers such that ${L < M < N}$, and
\begin{equation}
\phi_{n} \equiv \frac{\,\, n \, -L}{M-L}
\, .
\end{equation}
Note that in some sense the above definition incorporates both previous trials for $w_n$, either constant or equal to $v_n(x_*)$, along with a convex combination thereof. 
For reasons that will be clear in the following, we also need to assume that $L$ and $M$ depend on $N$ in such a way that, for ${N \to \infty}$, one has ${L \to \infty}$ and ${M \to \infty}$, yet ${L/M \to 0}$ and ${M/N \to 0}$.
Taking into account \eqref{eq:Gershgorin-type_bound_approx} and \eqref{eq:definizione_Delta}, we see that \eqref{eq:definizione_w_per_lowerbound} naturally suggests to split the maximum operation over different subsets, in formulae 
\begin{equation}
\max_{\vphantom{|} n=1}^{\vphantom{|} N} \Delta_{n} 
= 
\max \left\{
\max_{\vphantom{|} n=1}^{\vphantom{|} L} \Delta_{n} , 
\, \max_{\vphantom{|} n=M}^{\vphantom{|} N} \Delta_{n} , 
\max_{\vphantom{|} n=L+1}^{\vphantom{|} M-1} \Delta_{n} 
\right\}
\, .
\label{eq:Delta_max_composito}
\end{equation}
Let us discuss each different subset below.

\begin{enumerate}

\item For ${n=1,\dots,L}$, the fact that ${L/N \to 0}$ allows us to approximate ${J_n \approx \vartheta \oneminus{\vartheta}}$ (left-boundary current), so that from \eqref{eq:definizione_Delta} we can write
\begin{equation}
\Delta_{n} 
\approx 
\ddlim_{n+1} - \ddlim_{n-1}
+ \sqrt{\vphantom{\beta} \vartheta \oneminus{\vartheta}} \, 
\frac{{w}_{n+1} + {w}_{n-1}}{{w}_{n}} 
\qquad n = 1,\dots,L
\, .
\label{eq:Delta_subset1}
\end{equation}
Moreover, according to \eqref{eq:definizione_w_per_lowerbound} we have
\begin{subequations} 
\begin{eqnarray}
\frac{{w}_{n+1} + {w}_{n-1}}{{w}_{n}} 
& = 
\frac{{v}_{n+1} + {v}_{n-1}}{{v}_{n}} 
\qquad n = 1,\dots,L-1
\, ,
\label{eq:wratio_subset1_bulk}
\\
\frac{{w}_{L+1} + {w}_{L-1}}{{w}_{L}} 
& = 
\frac{{v}_{L+1} + {v}_{L-1}}{{v}_{L}}
+ \frac{1 - {v}_{L+1}/{v}_{L}}{M - L} 
\, .
\label{eq:wratio_subset1_boundary}
\end{eqnarray}
\end{subequations} 
Since ${L \to \infty}$, according to \eqref{eq:limite_rapporto_v} the ratio ${v}_{L+1}/{v}_{L}$ tends to a finite quantity, whereas ${M-L \to \infty}$, so that the last term in \eqref{eq:wratio_subset1_boundary} can be neglected, i.e.~\eqref{eq:wratio_subset1_bulk} can be used even for ${n=L}$. 
As a consequence, \eqref{eq:Delta_subset1} along with \eqref{eq:differenza_s-s} yields 
\begin{equation}
\Delta_{n} 
\approx 
2 {x}_{*} \sqrt{ \vphantom{\beta} \vartheta \oneminus{\vartheta} } 
\qquad n = 1,\dots,L
\, .
\label{eq:Delta_subset1_uguaglianza}
\end{equation}

\item For ${n=M,\dots,N}$, the fact that ${M \to \infty}$ allows us to neglect the exponentially decaying term ${\ddlim_{n+1} - \ddlim_{n-1}}$, so that from \eqref{eq:definizione_Delta} we can write 
\begin{equation}
\Delta_{n} 
\approx 
\sqrt{{J}_{n}} \, \frac{{w}_{n+1} + {w}_{n-1}}{{w}_{n}} 
\qquad n = M,\dots,N
\, .
\label{eq:Delta_subset2}
\end{equation}
Moreover, according to \eqref{eq:definizione_w_per_lowerbound} we have
\begin{subequations} 
\begin{eqnarray}
\frac{{w}_{n+1} + {w}_{n-1}}{{w}_{n}} 
= 2
\qquad \quad n = M+1,\dots,N
\, ,
\label{eq:wratio_subset2_bulk}
\\
\frac{{w}_{M+1} + {w}_{M-1}}{{w}_{M}} 
= 2 - \frac{1 - {v}_{M-1}/{v}_{L}}{M - L}
\, .
\label{eq:wratio_subset2_boundary}
\end{eqnarray}
\end{subequations} 
Since ${L,M \to \infty}$ and ${L/M \to 0}$, still by \eqref{eq:limite_rapporto_v} we can argue that ${{v}_{M-1}/{v}_{L} \to 0}$, whereas ${M-L \to \infty}$, so that the last term in \eqref{eq:wratio_subset2_boundary} can be neglected, i.e.~\eqref{eq:wratio_subset2_bulk} can be used for ${n = M}$ as well. 
As a consequence, \eqref{eq:Delta_subset2} yields 
\begin{equation}
\Delta_{n} 
\approx 
2 \sqrt{{J}_{n}} 
\qquad n = M,\dots,N
\, .
\end{equation}
Expressing $J_n$ by \eqref{eq:current_discrete-continuous} and taking into account that ${M/N \to 0}$, we obtain 
\begin{equation}
\max_{\vphantom{|} n=M}^{\vphantom{|} N} \Delta_{n} 
\approx 
2 \max \left\{ \!
\sqrt{ \vphantom{\beta} \vartheta \oneminus{\vartheta} } , 
\sqrt{ \beta \oneminus{\beta} }
\, \right\}
\, .
\label{eq:Delta_subset2_uguaglianza}
\end{equation}

\item For ${n=L+1,\dots,M-1}$, the fact that ${M/N \to 0}$ allows us to approximate ${J_n \approx \vartheta \oneminus{\vartheta}}$ (left-boundary current), whereas ${L \to \infty}$ also allows us to neglect the exponentially decaying term ${\ddlim_{n+1} - \ddlim_{n-1}}$. 
From \eqref{eq:definizione_Delta} we can then write 
\begin{equation}
\Delta_{n} 
\approx 
\sqrt{\vphantom{\beta} \vartheta \oneminus{\vartheta}} \, 
\frac{{w}_{n+1} + {w}_{n-1}}{{w}_{n}} 
\qquad n=L+1,\dots,M-1 
\, .
\label{eq:Delta_subset3}
\end{equation}
Moreover, according to \eqref{eq:definizione_w_per_lowerbound} we have
\begin{equation}
\frac{{w}_{n+1} + {w}_{n-1}}{{w}_{n}} 
= 
\frac{2 \phi_{n} {v}_{L} 
+ \oneminus{\phi_{n}} ({v}_{n+1} + {v}_{n-1})
+ \displaystyle \frac{{v}_{n-1} - {v}_{n+1}}{M-L}}
{\phi_{n} {v}_{L} + \oneminus{\phi_{n}} {v}_{n}} 
\end{equation}
for all $n$ in the range of interest. 
Let us now consider the three numerator terms in the right-hand side of this last equation, where we note that all displayed variables are positive. 
As far as the first term is concerned, we simply use the fact that ${x_* > 1}$ to state ${2 \phi_n v_L < 2 x_* \phi_n v_L}$. 
Regarding the second term, we take into account \eqref{eq:disuguaglianza_rapp-v}, which obviously implies ${\oneminus{\phi_{n}}({v}_{n+1} + {v}_{n-1}) < 2 x_* \oneminus{\phi_{n}} {v}_{n}}$. 
Regarding the third term, \eqref{eq:disuguaglianza_rapp-v} together with ${{v}_{n+1} > 0}$ immediately give also ${{v}_{n-1} - {v}_{n+1} < 2 x_* {v}_{n}}$, whereas we can use \eqref{eq:limite_rapporto_v} (i.e.~the fact that $v_n$ is eventually decreasing) to state ${\phi_{n} {v}_{L} + \oneminus{\phi_{n}} {v}_{n} > \phi_{n} {v}_{n} + \oneminus{\phi_{n}} {v}_{n} = {v}_{n}}$ for all ${n > L}$. 
In the end we obtain
\begin{equation}
\frac{{w}_{n+1} + {w}_{n-1}}{{w}_{n}} 
< 
2 x_* + \frac{2x_*}{M-L} 
\qquad n=L+1,\dots,M-1 
\, .
\end{equation}
Plugging this inequality into \eqref{eq:Delta_subset3}, we finally get 
\begin{equation}
\Delta_{n} 
\lessapprox 
2 x_* \sqrt{ \vphantom{\beta} \vartheta \oneminus{\vartheta} } 
\qquad n=L+1,\dots,M-1 
\, .
\label{eq:Delta_subset3_disuguaglianza}
\end{equation}

\end{enumerate}
We can now put together \eqref{eq:Delta_subset1_uguaglianza}, \eqref{eq:Delta_subset2_uguaglianza}, \eqref{eq:Delta_subset3_disuguaglianza} into \eqref{eq:Delta_max_composito}, and thence into \eqref{eq:Gershgorin-type_bound_approx}.
In the limit ${N \to \infty}$ we obtain
\begin{equation}
{\lambda_{\min}}^{(\infty)} 
\ge 
1 - 2 \max \left\{ {x}_{*} 
\sqrt{ \vphantom{\beta} \vartheta \oneminus{\vartheta} } , 
\sqrt{ \beta \oneminus{\beta} }
\, \right\}
\, .
\end{equation}
Comparing this last inequality with \eqref{eq:upperbound_all}, we easily conclude that the bound is tight. 
In the next (and last) subsection we give the proof that we had previously skipped.

\subsection{Proof of statement \eqref{eq:limite_rapporto_v}}

Let us first define 
\begin{equation}
  \z_{n} \equiv \frac{{v}_{n+1}(x_*)}{{v}_{n}(x_*)} 
  \qquad n = 1,2,\dots
  \, .
\end{equation}
Since we know that $v_n(x_*)$ is strictly positive for all ${n > 0}$, the definition is feasible and it also entails that $\z_{n}$ is itself strictly positive for all $n$. 
From \eqref{eq:definizione_v} we have 
\begin{subequations} 
\begin{eqnarray}
  \z_{1}   & = 2 x_* 
  - \frac{\ddlim_{2} - \ddlim_{0}}{\sqrt{\vartheta \oneminus{\vartheta}}} 
  \, , \\
  \z_{n+1} & = 2 x_* 
  - \frac{\ddlim_{n+2} - \ddlim_{n}}{\sqrt{\vartheta \oneminus{\vartheta}}} 
  - \frac{1}{\z_{n}}
  \qquad n = 1,2,\dots 
  \, .
  \label{eq:h_ricorsiva}
\end{eqnarray}
\end{subequations} 
The proof can be reduced to showing that, given two arbitrarily small positive real numbers $\varepsilon$ and $\delta$, the inequalities 
\begin{equation} 
\zeta(x_*) - \varepsilon < \z_{n} < \zeta(x_* - \delta)
\end{equation} 
are verified at least for $n$ large enough.
Thereafter, the thesis \eqref{eq:limite_rapporto_v} is easily achieved by sending $n$ to infinity and taking into account the continuity of $\zeta(x)$. 
In the following we report separate proofs for the lower and upper bounds.

\paragraph{Lower-bound} 

Let us consider any positive real number $\varepsilon$ such that
\begin{equation} 
\varepsilon < \zeta(x_*) 
\, ,
\label{eq:condizione_su_epsilon}
\end{equation} 
where the latter condition can be satisfied because ${\zeta(x_*) > 0}$ by construction. 
In the current hypotheses (entailing in particular ${\alpha < \vartheta}$), we know that $\ddlim_{n}$ is a monotonically increasing sequence, and therefore
\begin{equation} 
\frac{\ddlim_{n+2} - \ddlim_{n}}{\sqrt{\vartheta \oneminus{\vartheta}}}
> 0
\label{eq:deltas_gt_zero}
\end{equation} 
for all ${n~(\geq 0)}$.
From \eqref{eq:h_ricorsiva} we then have 
\begin{equation} 
\z_{n+1} < 2 x_* 
- \frac{1}{\z_{n}}
\, ,
\label{eq:disuguaglianza_L1} 
\end{equation} 
still for all ${n~(\geq 1)}$.
Taking into account \eqref{eq:definizione_z}, it is also possible to verify that, for any positive real number $\z$, the following inequality holds 
\begin{equation} 
2 x_* - \frac{1}{\z} \leq \zeta(x_*) + \frac{\z - \zeta(x_*)}{\zeta(x_*)^{2}} 
\, .
\label{eq:disuguaglianza_L2} 
\end{equation} 
Then, since $\z_{n}$ is positive by construction, from \eqref{eq:disuguaglianza_L1} and \eqref{eq:disuguaglianza_L2} we get
\begin{equation} 
\z_{n+1} < \zeta(x_*) + \frac{\z_{n} - \zeta(x_*)}{\zeta(x_*)^{2}} 
\, .
\label{eq:disuguaglianza_L3} 
\end{equation} 

Reasoning by contradiction, let us now assume that there exists some ${n~(\geq 1)}$ such that ${\z_{n} \leq \zeta(x_*) - \varepsilon}$, where \eqref{eq:condizione_su_epsilon} allows $\z_{n}$ to be positive. 
Then, by a repeated use of \eqref{eq:disuguaglianza_L3} we could argue that  
\begin{equation} 
\z_{n+k} < \zeta(x_*) - \frac{\varepsilon}{\zeta(x_*)^{2k}} 
\label{eq:disuguaglianza_L4} 
\end{equation} 
for any positive integer $k$, and therefore ${\z_{n+k} < 0}$ for some $k$, which is a contradiction. 
We can thus conclude that for any positive $\varepsilon$ satisfying \eqref{eq:condizione_su_epsilon} (and hence in particular for $\varepsilon$ arbitrarily close to $0$), the inequality 
\begin{equation} 
\z_{n} > \zeta(x_*) - \varepsilon 
\end{equation} 
must hold for all ${n~(\geq 1)}$.

\paragraph{Upper-bound}

Let us consider any positive real number $\delta$ such that
\begin{equation} 
\delta < x_* - 1
\, ,
\label{eq:condizione_su_delta}
\end{equation} 
where the latter condition can be satisfied because ${x_* > 1}$ by hypothesis.
We know that the term ${\ddlim_{n+2} - \ddlim_{n}}$ decays exponentially for ${n \to \infty}$ and, as a consequence, there exists some integer $m$ (depending on $\delta$) such that 
\begin{equation} 
\frac{\ddlim_{n+2} - \ddlim_{n}}{\sqrt{\vartheta \oneminus{\vartheta}}}
< \delta
\label{eq:deltas_lt_delta}
\end{equation} 
for all ${n \geq m}$. 
From \eqref{eq:h_ricorsiva} we then have 
\begin{equation} 
\z_{n+1} > 2 x_* 
- \frac{1}{\z_{n}}
- \delta 
\, ,
\label{eq:disuguaglianza_U1} 
\end{equation} 
still for all ${n \geq m}$. 
Using this last inequality, let us first prove that ${\z_{n} < 1}$ for all ${n \geq m}$. 
Reasoning by contradiction, if we had ${\z_{n} \ge 1}$ for some ${n \geq m}$, then \eqref{eq:disuguaglianza_U1} and \eqref{eq:condizione_su_delta} with ${x_* > 1}$ would imply ${\z_{n+1} > 1}$, and therefore by induction $\z_{n}$ eventually larger than $1$. 
This is clearly in contradiction with the fact that $v_n(x_*)$ is bounded (up to a constant prefactor) by the exponentially decreasing sequence ${\zeta(x_*)}^{n}$.  
Now, taking into account \eqref{eq:definizione_z}, it is possible to verify that, given a positive real number ${\z < 1}$, the inequality 
\begin{equation} 
2 x_* - \frac{1}{\z} \geq \z + 2 \delta 
\label{eq:disuguaglianza_U2} 
\end{equation} 
is verified if and only if
\begin{equation} 
\z \geq \zeta(x_* - \delta)
\, ,
\label{eq:disuguaglianza_U3} 
\end{equation} 
where \eqref{eq:condizione_su_delta} ensures that $\zeta(x_* - \delta)$ is real (positive and less than $1$).

Again reasoning by contradiction, let us now assume that there exists some ${n \geq m}$ such that ${\z_n \geq \zeta(x_* - \delta)}$. 
Then, since it must be ${\z_{n} < 1}$ because of the above argument, \eqref{eq:disuguaglianza_U1} and the equivalence between \eqref{eq:disuguaglianza_U2} and \eqref{eq:disuguaglianza_U3} imply the inequality
\begin{equation} 
\z_{n+1} > \z_{n} + \delta 
\, ,
\label{eq:disuguaglianza_U4} 
\end{equation} 
also entailing ${\z_{n+1} > \zeta(x_* - \delta)}$. 
Proceeding by induction, we obtain ${\z_{n+k} > 1}$ for some positive integer $k$, which is clearly a contradiction. 
We can thus conclude that for any positive $\delta$ satisfying \eqref{eq:condizione_su_delta} (and hence in particular for $\delta$ arbitrarily close to $0$), the inequality 
\begin{equation} 
\z_{n} < \zeta(x_* - \delta)
\end{equation} 
must hold for large enough $n$.

\end{document}